\documentclass[10pt,journal]{IEEEtran}

\usepackage{blindtext, graphicx}
\usepackage{adjustbox}
\usepackage{algorithmic}
\usepackage{algorithm}
\usepackage[english]{babel}
\usepackage{amsmath}
\usepackage{amssymb}
\usepackage{enumitem}
\usepackage{epstopdf}
\usepackage{subcaption}
\usepackage{overpic}
\def\cput(#1,#2)#3{\put(#1,#2){\hbox to 0pt{\hss{#3}\hss}}}

\newtheorem{lemma}{\textbf{Lemma}}

\ifCLASSINFOpdf
\else
\fi

\begin{document}
%
\title{$n$-VDD: Location Privacy Protection Based on Voronoi-Delaunay Duality}



%

\author{Wei Zeng$^\ddagger$, 
Abdur B. Shahid,
Keyan Zolfaghari,
Aditya Shetty,
Niki Pissinou, 
and~Sitharama S. Iyengar 
\thanks{W. Zeng, A. B. Shahid, N. Pissinou and S. S. Iyengar were with the School of Computing and Information Sciences,
Florida International University, Miami, FL 33199, USA. K. Zolfaghari was with University of Miami, Miami, FL 33124, USA. A. Shetty was with University of North Carolina at Chapel Hill, NC 27515, USA.
$^\ddagger$Corresponding e-mail: wzeng@cs.fiu.edu.}
}

\maketitle

\begin{abstract}

To date, location privacy protection is a critical issue in Location-Based Services (LBS). In this work, we propose a novel geometric framework based on the classical discrete geometric structure, the Voronoi-Delaunay duality (VDD). We utilize the fact that the user location cannot be recovered if only given an irregular $n$-sided Voronoi cell around it, and the anonymity zone is the intersection of all the parallel strips perpendicular to and bounded by $n$ Voronoi edges. The irregular Voronoi cell and its variations can be used as the concealing space to hide the user location or the region of interest and submitted to the LBS server. Within this framework, we propose multiple typical anonymizing models by introducing irregularity to the convex regular VDD structure by shifting the interior Voronoi cell, exterior Delaunay polygon, sector rays, or their combinations. The proposed methods are efficient by taking advantage of the VDD principle where main computations are linear line-line intersections. Experiments with various parameters demonstrate the efficiency and efficacy of the proposed $n$-VDD framework.

\end{abstract}


%
\IEEEpeerreviewmaketitle

\section{Introduction}

With the help of pervasive global positioning system (GPS) and radio frequency identification (RFID) enabled mobile computing, the market share of location-based services (LBS) is increasing rapidly \cite{first}. Today, people frequently use LBS to find the nearest ATM, restaurant, hospital, or gas station. Social networks, including but not limited to Facebook, LinkedIn and Twitter, as well as Internet telephony service providers such as Skype have all created frameworks for geosocial networking. LBS providers require users to divulge their exact location but guarantee a higher quality of service (QoS) with more accurate location data.
If one uses Yelp to look for restaurants in a 15 mile radius, it would not matter if the location data is not as accurate, with the search being conducted in a relatively large area. Conversely, if a user looks to find the nearest gas station, Google Map would require a more accurate location to ensure a higher QoS. A user may enjoy receiving services with higher precisions, but is unaware of the possible exploitations of his/her location data. LBS providers can use this data to understand a user's mobility pattern, enabling them to send unwanted advertisements;  
and attackers can perform malicious attacks. 
This has led to a general consensus that protecting a user's location privacy is a highly important issue in LBS.

Previous studies have designed location privacy models based on a trade-off between location privacy and QoS. If the degree of privacy is high, then the QoS is low and vice versa. So, it is important to understand this trade-off while designing a LBS privacy model. 
We classify the previous methods to protect location privacy into three main categories \cite{Wernke}: 1) $k$-anonymity, 2) position dummies, and 3) spatial obfuscation.

In \emph{$k$-anonymity}, the framework pairs a user's location with $k$-1 locations of other neighboring users \cite{Gruteser,gedik2005location,zhang2015real,kido2005anonymous,gong2010protecting,sweeney2002achieving,machanavajjhala2007diversity} \cite{gedik2004customizable,lefevre2006mondrian,el2008protecting} and provides the LBS provider with a box containing that user and the other $k-1$ locations. This approach engenders several problems. It has an inherent dependency on the presence of the other $k-1$ users in the specified region or time. Furthermore, each user cannot have a personalized privacy setting with the settings of the other neighboring users overlapping. Gedik et al. proposed the CliqueCloak theorem \cite{gedik2008protecting} to improve the original $k$-anonymity approach that joins multiple queries that overlap together into a clique, and sends the minimum bounding rectangle of those users as a single query. Although it is advantageous in the case of reducing computation time and increasing the privacy of the users, it can fail to join users into a clique. Marius et al. \cite{Wernke} classified several improved methods of $k$-anonymity: strong $k$-anonymity \cite{zhang2009cloaking,Talukder2010}, l-diversity \cite{Bamba2008}, $t$-closeness \cite{4221659}, $p$-sensitivity\cite{Solanas2008}, and historical $k$-anonymity \cite{5088932,xiao2006personalized,aggarwal2008general,li2007t}. In these schemes, the probability of identifying a user is $1/k$. These approaches also suffer from a high communication and query processing cost \cite{fung2010privacy}.

The \emph{position dummy} based techniques direct multiple fake positions along with user's true location to location service providers \cite{Duckham:2005:FMO:2154273.2154286,Lu:2008:PPA:1626536.1626540,Suzuki:2010:ULA:1869790.1869846, kato2012dummy, tran2010binomial,kato2013user}. Hua et al. \cite{Lu:2008:PPA:1626536.1626540} proposed a dummy-based method, titled PAD, which generates dummy users in a virtual grid or circle. Qilang et al. \cite{Suzuki:2010:ULA:1869790.1869846} proposed a user dummy generation based clustering method to provide privacy on road networks. Location obfuscation methods have the advantage of generating dummy locations by itself, unlike the $k$-anonymity based methods. However, generating indistinguishable dummy locations is a challenge for these methods \cite{Wernke}.

To address the problems of $k$-anonymity and dummy based methods, some solutions propose \emph{spatial obfuscation} methods \cite{Ardagna:2007:LPP:1770560.1770566,Gutscher:2006:CTS:1898699.1898907,6567113, 6808227,kalnis2007preventing,mokbel2006new,chow2006peer,ardagna2011obfuscation,bamba2008supporting}, based on the idea of providing a user-defined obfuscation area without revealing explicit location information at the expense of quality of services. Ardagna et al. \cite{Ardagna:2007:LPP:1770560.1770566} proposed a method to submit a circle instead of a user's exact location. Different methods were presented to use other geometric transformations to preserve a user's location. Kalnis et al. \cite{} proposed  transformations based on the $k$-anonymity concept for nearest neighbor search, while hiding user's location.  Gutscher et al. \cite{Gutscher:2006:CTS:1898699.1898907} detailed the use of coordinate transformation to protect a user's privacy. Min et al. \cite{6808227} proposed a privacy scheme based on a line-symmetric transformation 
for database privacy in cloud computing. Li et al. \cite{6567113} proposed a geometric approach towards location privacy that divides the user's region of interest (ROI) into $n$ concealing disks ($n$-CD) and submits the centers and the radii of those concealing disks for transmission. Guo et al. \cite{guo2015pseudonym} extended $n$-CD by introducing dynamic pseudonyms-changing mechanism with the expense of concealing and communication costs. The proposed framework in this work falls into this category, and is comparable to the $n$-CD approach.

\subsection{Our Approach}

\begin {table}[t]
\caption {Table of Symbols\label{tab:symbols}}
\begin{center}
\begin{tabular}
{ l | p{5cm}} \hline Symbol & Definition \\ \hline
$O$ & User's location (or seed)  \\ 
$n$ & Number of vertices of polygon around $O$\\ 
$\mathbf{P_c}$ ($\mathbf{P_c'}$/$\mathbf{P_c^*}$)& Delaunay polygon (shifted/scaled)  \\ 
$C_i$ ($C_i'$/$C_i^*$) & Delaunay polygon vertices (shifted/scaled) \\ 
$\mathbf{P_v}$ ($\mathbf{P_v'}$/$\mathbf{P_v^*}$) & Voronoi polygon (shifted/scaled) \\ 
$V_i$ ($V_i'$/$V_i^*$) & Voronoi polygon vertices (shifted/scaled) \\ 
$\mathbf{A_z}$ ($\mathbf{A_z'}$/$\mathbf{A_z^*}$) & Anonymity zone (shifted/scaled) \\ 
$A_i$ ($A_i^*$) & Anonymity zone vertices (scaled) \\ 
$\alpha$& $\frac{2\pi}{n}$, sector angle \\ 
$|\overline{OX}|$ & Length of line segment $\overline{OX}$  \\ 
$\tau$ & Range of vertices on a line  \\ 
$\kappa$ & Range of random angle adaption \\
$r$ & User-defined radius of interest  \\ 
$\lambda$ & Scaling factor  \\
$\Psi$ & Concealing cost\\
$\Gamma$& Privacy level\\ \hline
\end{tabular}
\end{center}
\end{table}

In this work, we propose a novel framework based on the classical geometric structure, the so-called \emph{Voronoi-Delaunay Duality} (VDD). Table \ref{tab:symbols} gives the symbols used in the work.

As shown in Fig. \ref{fig:VDD}, we treat the user location $O$ as the center and properly select $n$ discrete points $\mathbf{C}=\{C_0, C_1, ..., C_{n-1}\}$ ($n\geq 3$) on the Euclidean plane surrounding $O$ to result a Delaunay triangulation $\mathbf{T}$, by connecting edges $C_iO$ and $C_iC_{i+1}$ ($C_n=C_0$).
Each triangle $\vartriangle C_iOC_{i+1}$ satisfies empty circle criterion, i.e., the circumcircle of the triangle doesn't contain any other point.
The Voronoi diagram of $\mathbf{C}\cup\{O\}$ is computed by the perpendicular bisector lines of all the edges of $\mathbf{T}$ and their intersection points $\mathbf{V}=\{V_0, V_1, ..., V_{n-1}\}$. The Voronoi diagram and the Delaunay triangulation $\mathbf{T}$ are dual to each other. The dual relationship is unique.
The VDD gives a two-layered dual structure defined around the user location $O$, including the exterior convex \emph{Delaunay polygon} $\mathbf{P_c}=\langle C_0C_1...C_{n-1}\rangle$ and the interior convex \emph{Voronoi polygon} $\mathbf{P_v}=\langle V_0V_1...V_{n-1}\rangle$, as shown in Fig. \ref{fig:VDD}(a). The facts include:

1) $\mathbf{P_c}$ and $O$ define the interior $\mathbf{P_v}$ uniquely; 

2) $\mathbf{P_v}$ and $O$ define $\mathbf{P_c}$ uniquely; and

3) $\mathbf{P_c}$ and $\mathbf{P_v}$ define $O$ uniquely.

However, if only $\mathbf{P_v}$ and without any other information, one cannot figure out the exact location of the center $O$.
But it can give the feasible region containing the user location.
Based on the VDD perpendicular rule, we draw the parallel strip perpendicular to each edge of Voronoi cell, the intersecting region of all the parallel strips and the Voronoi cell is a convex polygon $\mathbf{A_z}=\langle A_0A_1...A_m \rangle $, where any point may be the user location. This region is the so-called \emph{anonymity zone} in our framework. Figure \ref{fig:VDD}(a) shows one example when $n=5$ where the anonymity zone $A_z$ is within the Voronoi cell, and (b) gives two examples when $n=3,4$, where the anonymity zone is the Voronoi polygon itself. It means that at a Voronoi polygon vertex with corner angle $\leq \pi/2$, the associated perpendicular lines are either outside the Voronoi cell or overlap the Voronoi edges, and then the associated Voronoi polygon edges are used for computing the intersection.
Therefore, the interior Voronoi polygon $\mathbf{P_v}$ can be used to hide the user location $O$.
In the $n$-VDD framework, we submit the Voronoi polygon or its variations to the LBS provider.

{\setlength{\tabcolsep}{0pt}
\begin{figure}[t]
\centering
\footnotesize
\begin{tabular}{cc}
\includegraphics[height=0.35\linewidth]{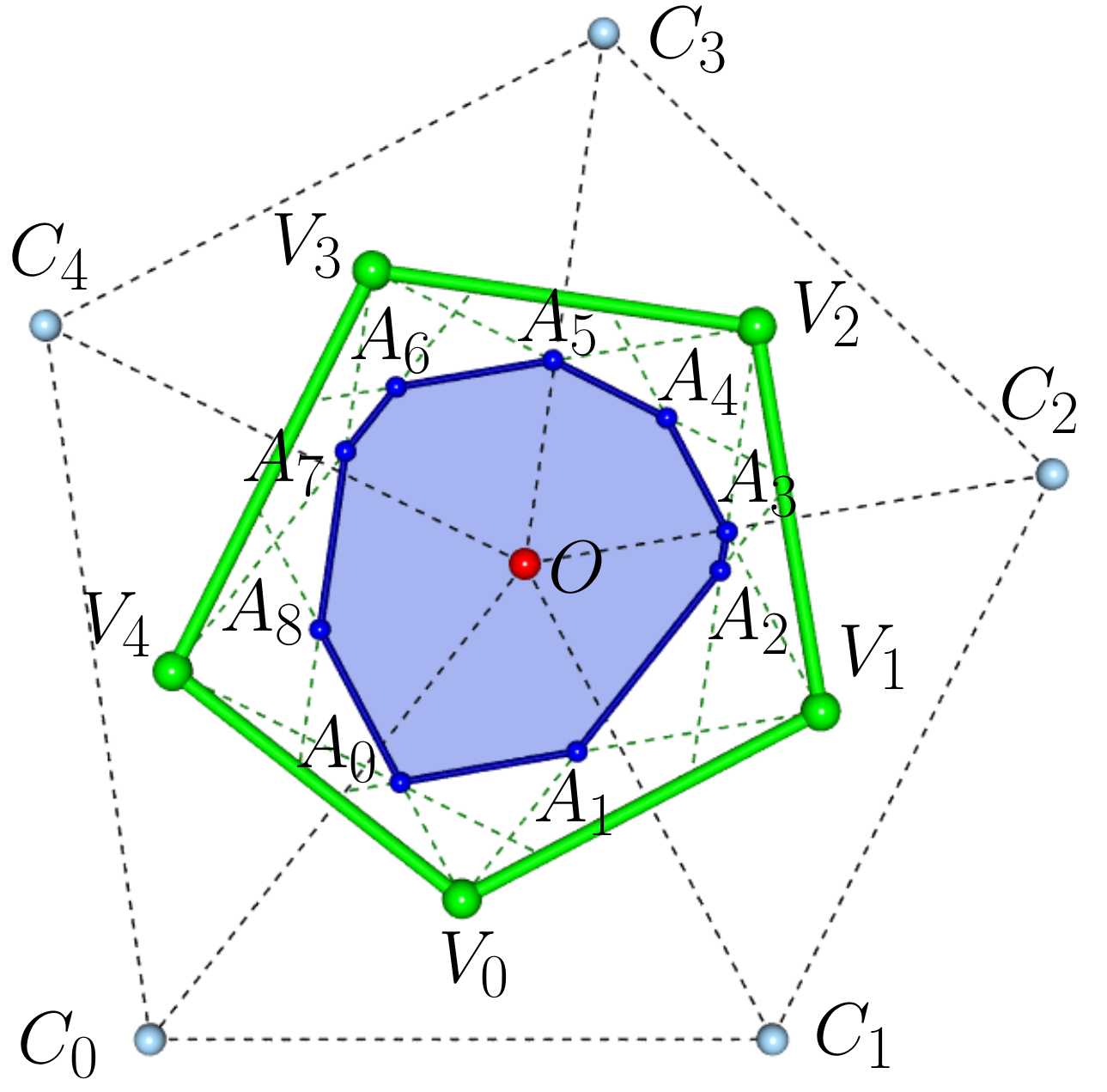}&
\includegraphics[height=0.3\linewidth]{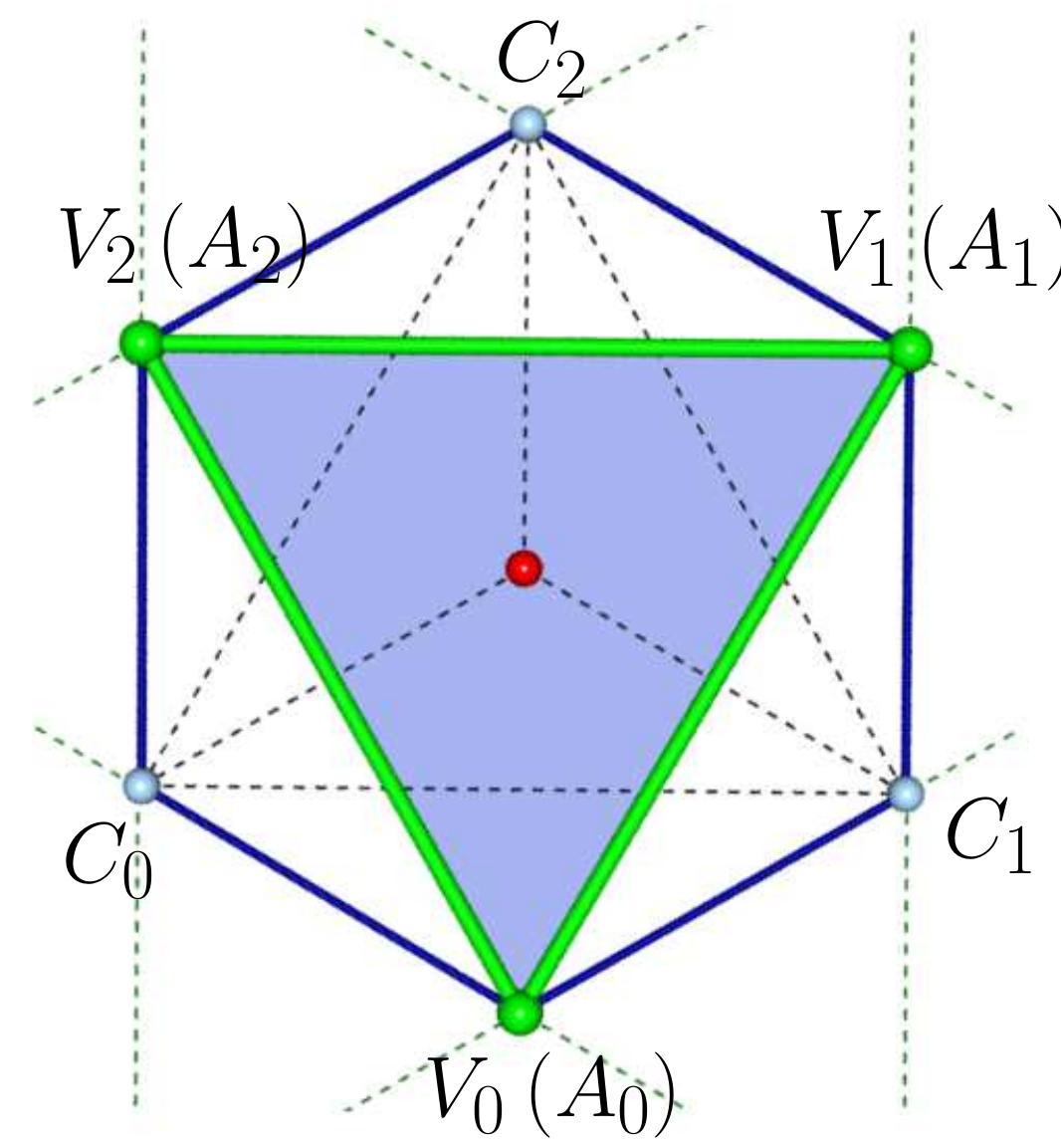}
\includegraphics[height=0.3\linewidth]{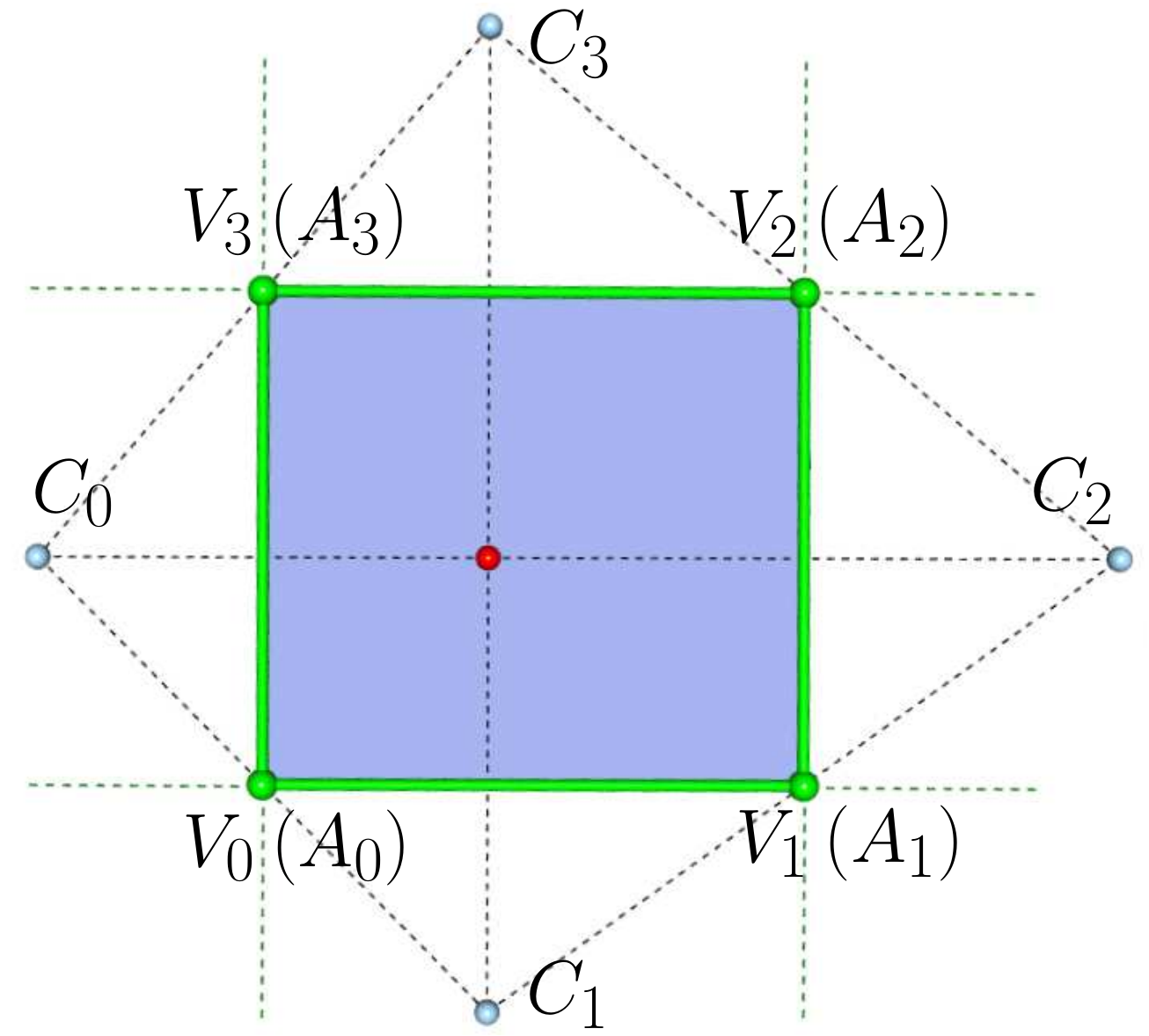}\\
(a) Anonymity zone & (b) Examples when $n=3,4$\\
from Voronoi polygon & with corner angles $\leq \pi/2$ \\
\end{tabular}
\caption{Voronoi-Delaunay duality and anonymity zone. \label{fig:VDD}}
\end{figure}
}

{\setlength{\tabcolsep}{0pt}
\begin{figure}[h]
\centering
\footnotesize
\begin{tabular}{ccc}
{\begin{overpic}[width=.33\linewidth]{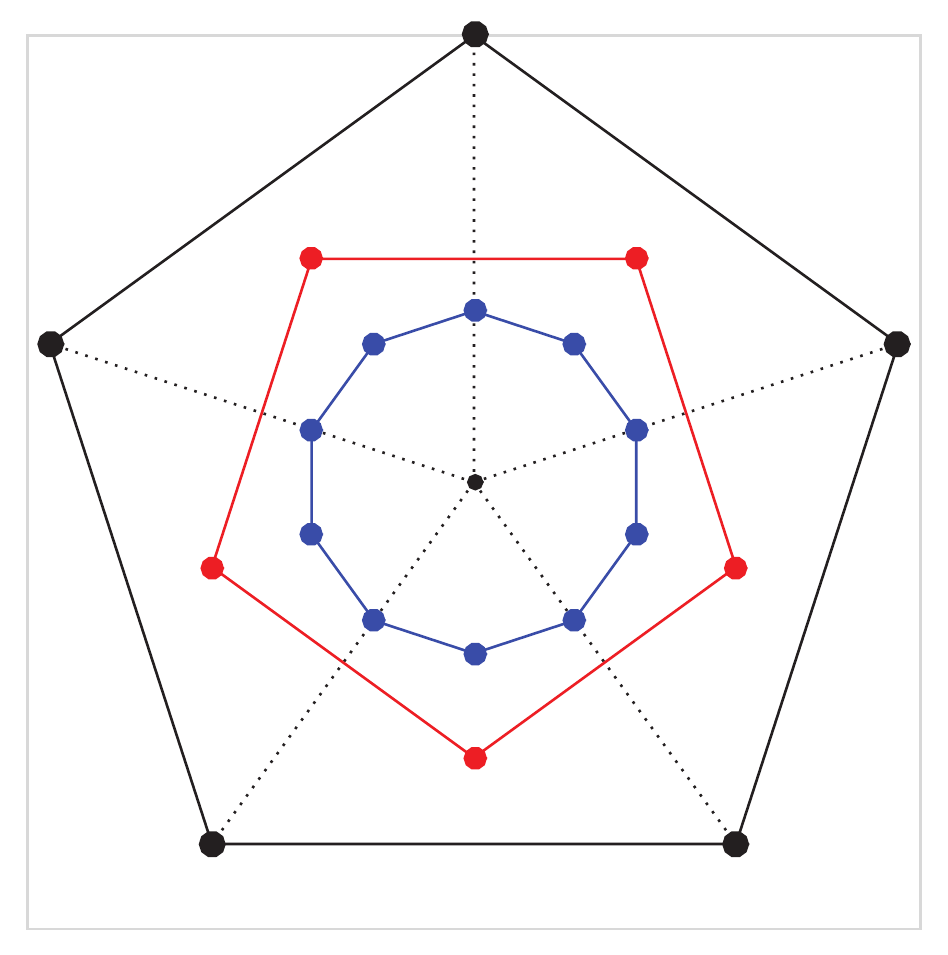}
\tiny
\cput(50,80){$\mathbf{T}$}
\cput(12,10){$\mathbf{P_c}$}
\cput(30,25){$\mathbf{P_v}$}
\cput(45,41){$\mathbf{A_z}$}
\cput(55,50){$O$}
\end{overpic}}&
\includegraphics[width=.33\linewidth]{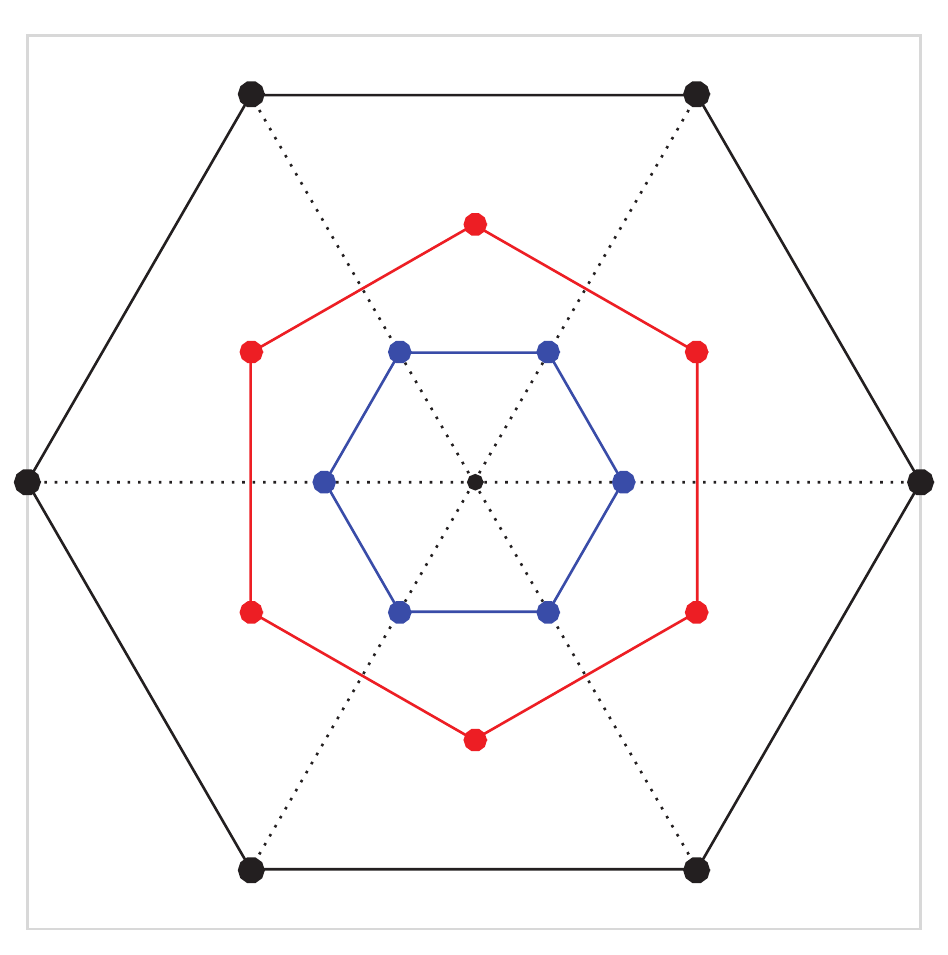}&
\includegraphics[width=.33\linewidth]{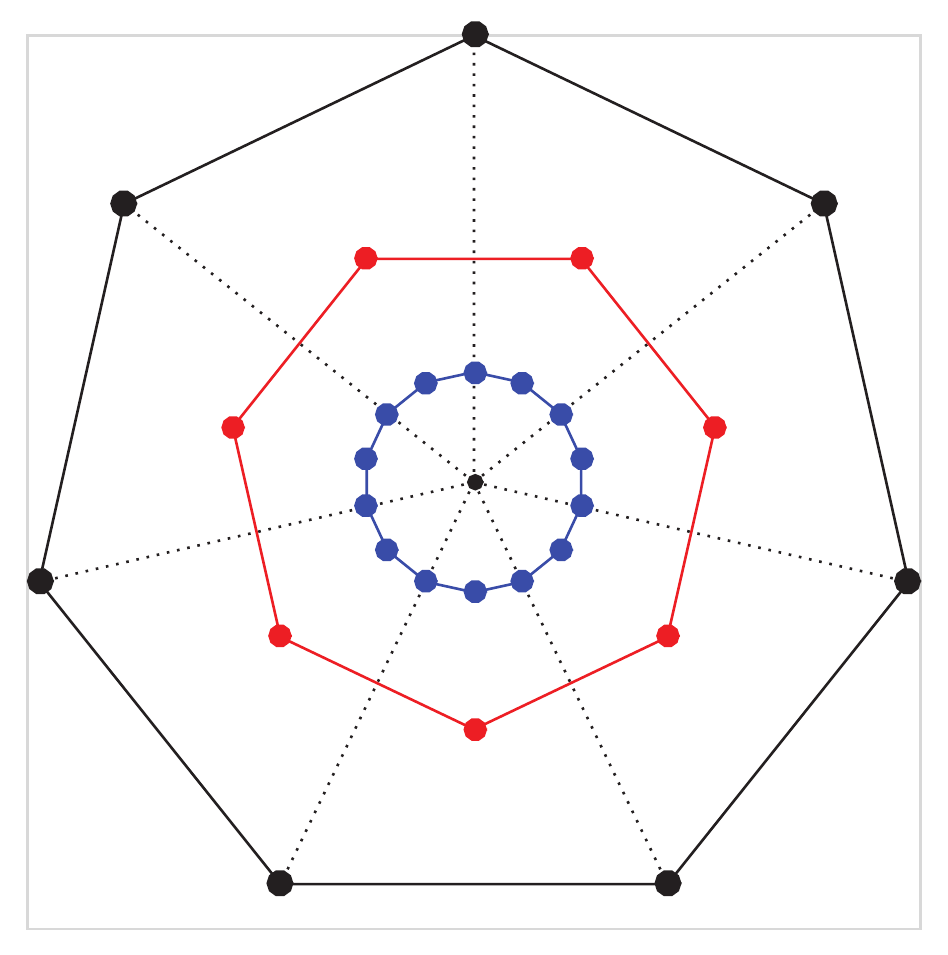}\\
\end{tabular}
\caption{VDD of convex regular polygons. \label{fig:VDD_2}}
\end{figure}
}

The unique degenerated case is that if $\mathbf{P_c}$ is a convex regular polygon\footnote{In Euclidean geometry, a regular polygon is a polygon that is equiangular and equilateral.} taking $O$ as the centroid, then the $\mathbf{P_v}$ is regular with the same centroid, as shown in Fig. \ref{fig:VDD_2}. In this case, if there is such a principle in the protocol public to users/attackers: $\mathbf{P_c}$ is regular around $O$, then $O$ can be easily recovered from $\mathbf{P_v}$, i.e., the centroid of $\mathbf{P_v}$. To conquer this, our strategy is to introduce irregularity to the VDD structure of a convex regular polygon, in detail, by shifting the exterior polygon $\mathbf{P_c}$, or interior polygon $\mathbf{P_v}$, or both, to be irregular. The following three models are given to demonstrate the performance of the $n$-VDD framework.

\begin{description}
  \item[Model I - Interior Shifting.] $\mathbf{P_c}$ is a regular $n$-sided convex polygon with the centroid at $O$. We adapt the corresponding Voronoi cell $\mathbf{P_v}$ to be irregular by parallelly shifting each Voronoi edge along the dual Delaunay edge while keeping all the Voronoi cell vertices within each original sector and $\mathbf{P_c}$. The shifting position is uniformly and randomly selected within a range calculated based on the above condition, which guarantees the user location $O$ is included in the shifted Voronoi cell.
  \item[Model II - Exterior Shifting.] We shift $\mathbf{P_c}$ to be an irregular $n$-sided convex polygon around $O$ such that the corresponding triangulation $\mathbf{T}$ is still Delaunay. The resulted Voronoi cell $\mathbf{P_v}$ is also irregular.
For each Delaunay vertex $C_{i+1}$, we compute the valid range on its sector ray such that the Voronoi vertex $V_i$ is within the corresponding sector $\sphericalangle C_iOC_{i+1}$, and then we randomly and uniformly select a point in the valid range as $C_{i+1}$.
  \item[Model III - Double Shifting.] We first shift $\mathbf{P_c}$ to be an irregular $n$-sided convex polygon around $O$ as Model II, to guarantee the corresponding triangulation $\mathbf{T}$ is Delaunay, and then we shift the resulted Voronoi cell $\mathbf{P_v}$ as Model I. This is a combination of Models I, II. Generally, both the exterior and interior polygons are irregular.

\end{description}

\subsubsection*{Sector Shifting} Another aspect to generate irregularity from the convex polygon is to shift the sector rays to make the sector angles not equal. We integrate the sector shifting operation into the above three Models I, II and III, to generate three variation Models I$_\alpha$, II$_\alpha$ and III$_\alpha$ accordingly. In each anonymizing model, we first randomly and uniformly perturb the sector rays in a range to achieve the sector angle inequality, and then apply other anonymizing principles.
Note that the randomness of sector rays itself on a convex regular polygon can generate an irregular convex polygon with the same sector radius. The resulted Voronoi cell is irregular. We call this Model $\alpha$. However, it can not be used for anonymization. That is because the center $O$ can be easily calculated by the intersection of the bisectors of two Voronoi edges due to the property of the same sector radius, as shown in Fig. \ref{fig:model_a} with a simulation example. Therefore, only angle randomness on the convex regular polygon is not enough for hiding the user location.

{\setlength{\tabcolsep}{0pt}
\begin{figure}[ht]
\centering
\footnotesize
\begin{tabular}{c}
\includegraphics[height=.45\linewidth]{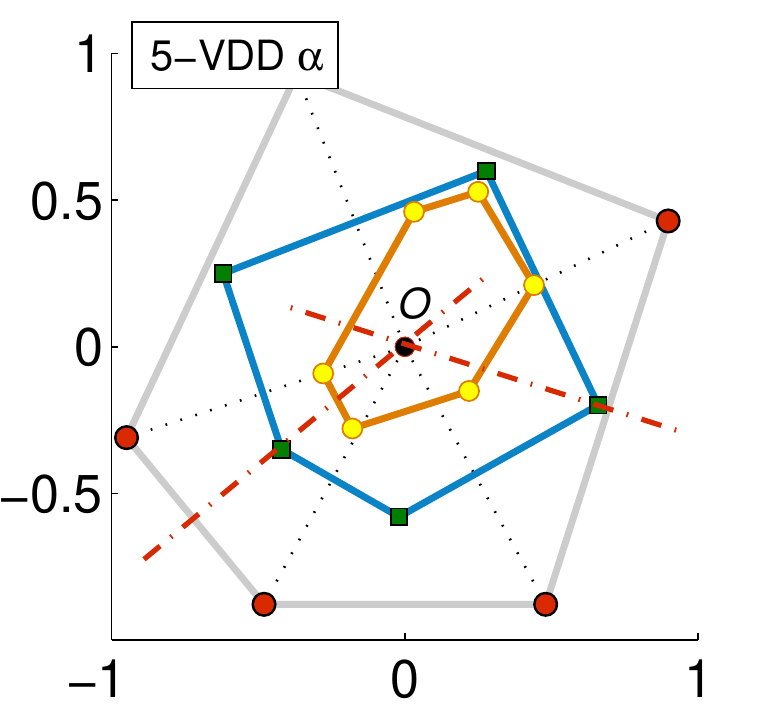}
\end{tabular}
\caption{Illustration of the VDD model with only angle randomness $\alpha$ on a convex regular polygon, which reveals the center $O$ and can not be used for anonymization. \label{fig:model_a}}
\end{figure}
}

In summary, there are three irregularity principles to adapt the convex regular polygon to achieve irregular VDD structures, (1) interior shifting, (2) exterior shifting and (3) sector shifting. The combinations of the three can form totally 7 models (see Table \ref{tab:models}), among which the Model $\alpha$ is not applicable for anonymization, and other 6 Models I, II, III, I$_\alpha$, II$_\alpha$, and III$_\alpha$ are to be used for anonymization (see Figs. \ref{fig:model_I}, \ref{fig:model_II} and \ref{fig:model_III}). Details will be explained in later sections.

{\setlength{\tabcolsep}{2pt}
\begin {table*}[t]
\caption {The $n$-VDD models. IR - Irregularity. Model $\alpha$ is not applicable for anonymization.}  \label{tab:models}
\centering
\begin{tabular}
{  c | c | c | c | c | c } \hline
~Model~ & ~IR.1 Interior Sifting~ & ~IR.2 Exterior Shifting~ & ~IR.3 Sector Shifting~ & ~Anonymizing Protocol~  & ~Generated Polygons~\\ \hline
I & $\surd$& -&   -     & \textsl{S0} $\rightarrow$ \textsl{R1} $\rightarrow$ \textsl{P1} $\rightarrow$ \textsl{R0}& $\mathbf{P_c}\rightarrow\mathbf{P_v}\rightarrow\mathbf{P_v'}\rightarrow\mathbf{P_v^*}$\\
II &- & $\surd$& -         & \textsl{S0} $\rightarrow$ \textsl{P2} $\rightarrow$ \textsl{R1} $\rightarrow$ \textsl{R0}&
$\mathbf{P_c}\rightarrow\mathbf{P_c'}\rightarrow\mathbf{P_v'}\rightarrow\mathbf{P_v^*}$\\
III & $\surd$ & $\surd$ &   -       & \textsl{S0} $\rightarrow$ \textsl{P2} $\rightarrow$ \textsl{R1} $\rightarrow$ \textsl{P1} $\rightarrow$ \textsl{R0}&
$\mathbf{P_c}\rightarrow\mathbf{P_c'}\rightarrow\mathbf{P_v'}\rightarrow\mathbf{P_v'}\rightarrow\mathbf{P_v^*}$\\ \hline
$\alpha$& - & - & $\surd$ &
N/A&
N/A\\
 \hline
I$_\alpha$  & $\surd$&- & $\surd$ & \textsl{S1} $\rightarrow$ \textsl{R1} $\rightarrow$ \textsl{P1} $\rightarrow$ \textsl{R0}& $\mathbf{P_c'}\rightarrow\mathbf{P_v'}\rightarrow\mathbf{P_v'}\rightarrow\mathbf{P_v^*}$\\
II$_\alpha$ &- &$\surd$ & $\surd$ &\textsl{S1} $\rightarrow$ \textsl{P2} $\rightarrow$ \textsl{R1} $\rightarrow$ \textsl{R0}&
$\mathbf{P_c'}\rightarrow\mathbf{P_c'}\rightarrow\mathbf{P_v'}\rightarrow\mathbf{P_v^*}$\\
III$_\alpha$&  $\surd$ & $\surd$ & $\surd$ &~\textsl{S1} $\rightarrow$ \textsl{P2} $\rightarrow$ \textsl{R1} $\rightarrow$ \textsl{P1} $\rightarrow$ \textsl{R0}~&
~$\mathbf{P_c'}\rightarrow\mathbf{P_c'}\rightarrow\mathbf{P_v'}\rightarrow\mathbf{P_v'}\rightarrow\mathbf{P_v^*}$~\\  \hline
\end{tabular}
\end{table*}
}

{\setlength{\tabcolsep}{0pt}
\begin{figure*}[ht]
\centering
\footnotesize
\begin{tabular}{cccc}
\includegraphics[width=.25\linewidth]{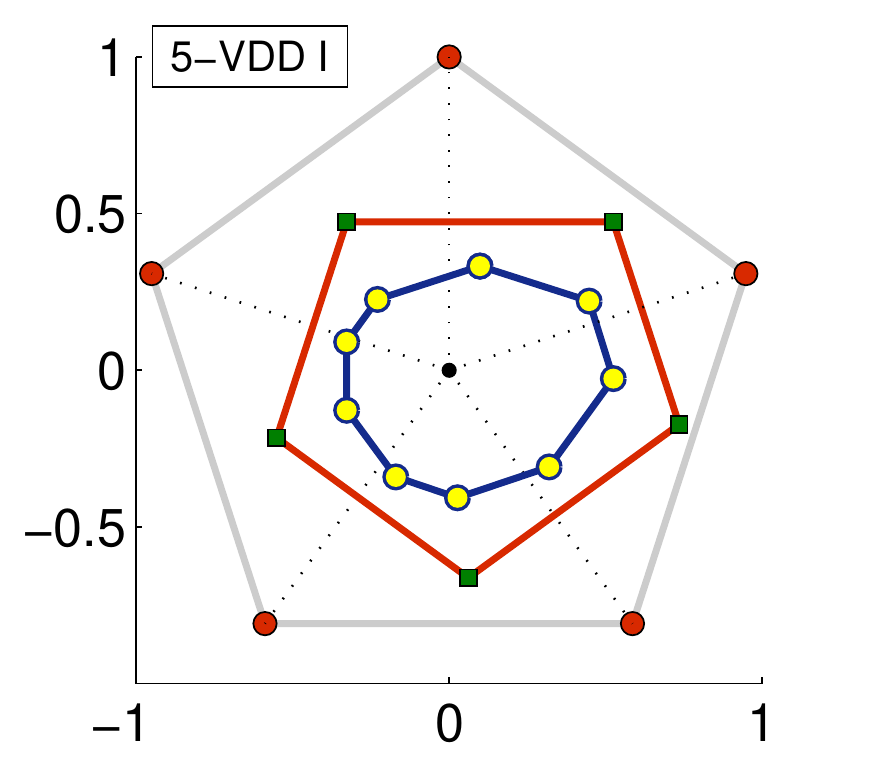}&
\includegraphics[width=.25\linewidth]{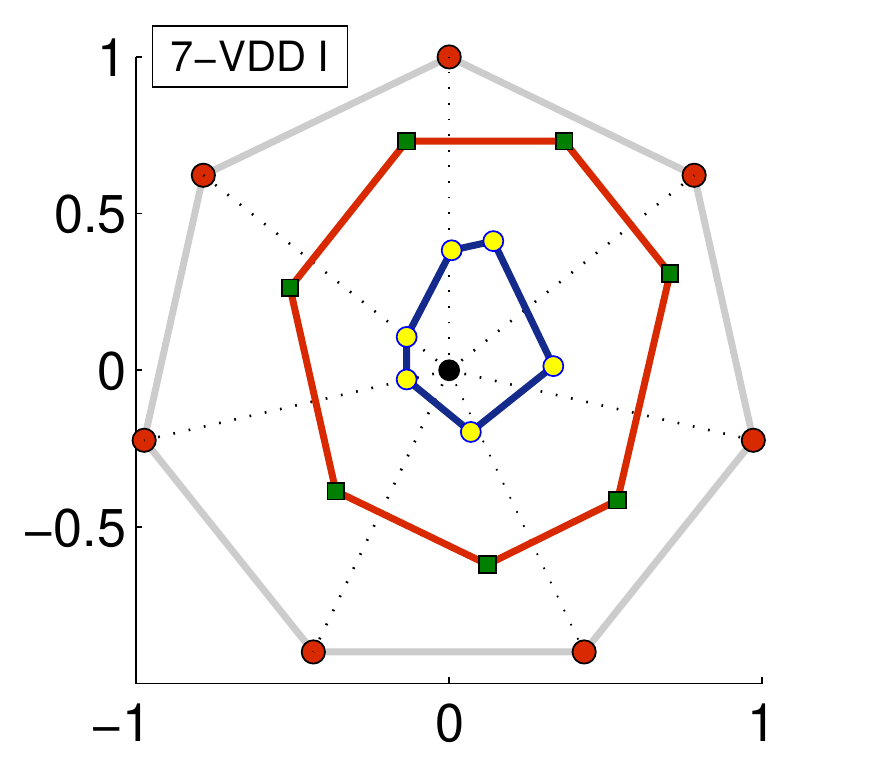}&
\includegraphics[width=.25\linewidth]{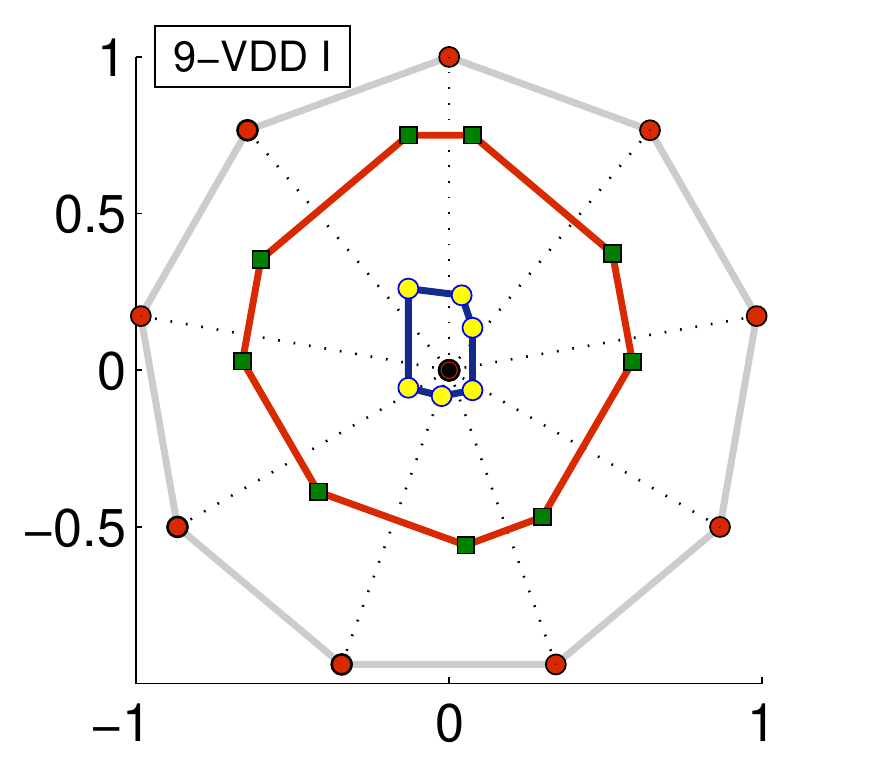}&
\includegraphics[width=.25\linewidth]{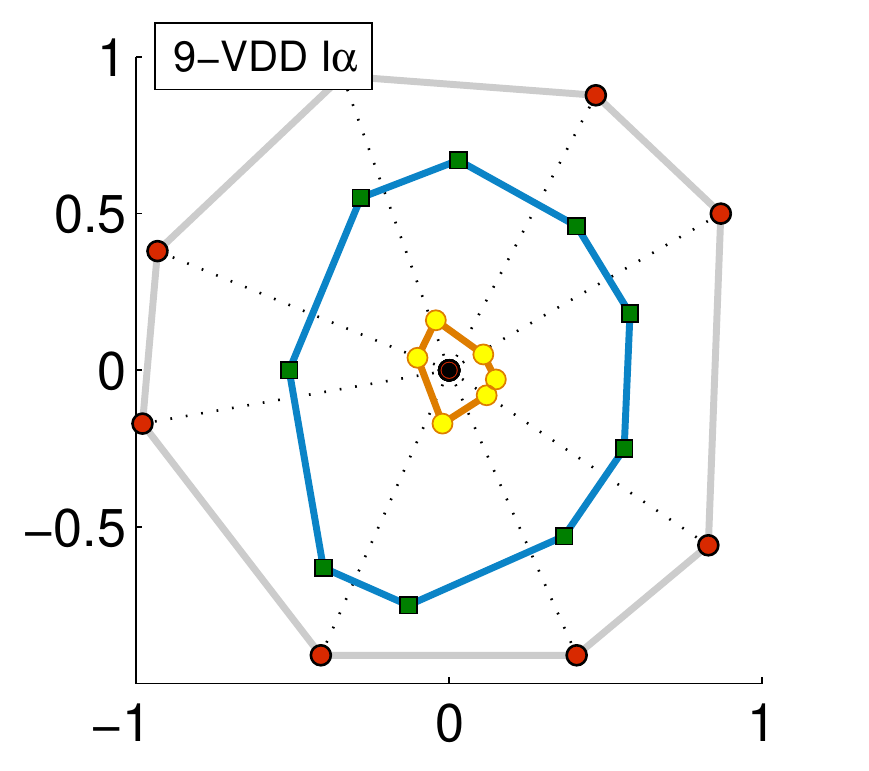}\\
\end{tabular}
\caption{Simulation examples of the $n$-VDD models I and I$_\alpha$ before scaling by ROI radius. \label{fig:model_I}}
\end{figure*}
}

{\setlength{\tabcolsep}{0pt}
\begin{figure*}[ht]
\centering
\footnotesize
\begin{tabular}{cccc}
\includegraphics[width=.25\linewidth]{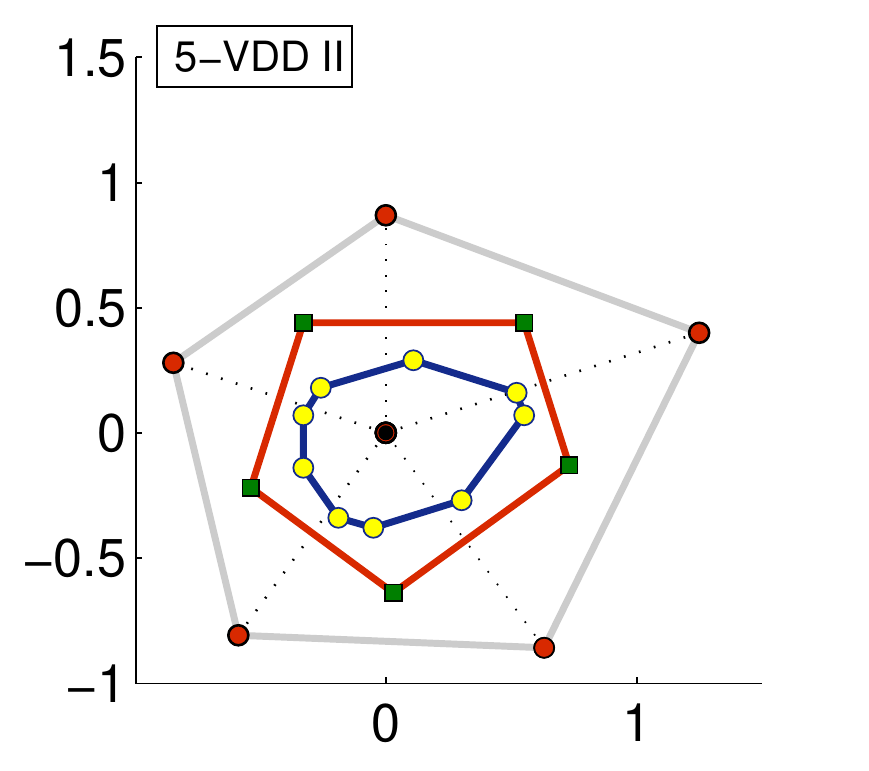}&
\includegraphics[width=.25\linewidth]{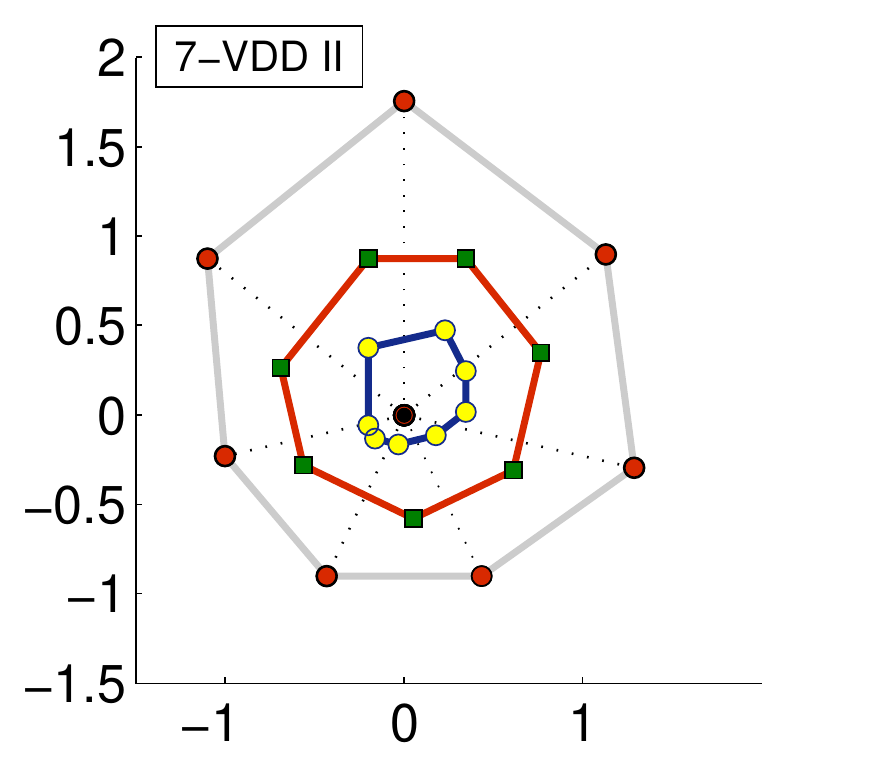}&
\includegraphics[width=.25\linewidth]{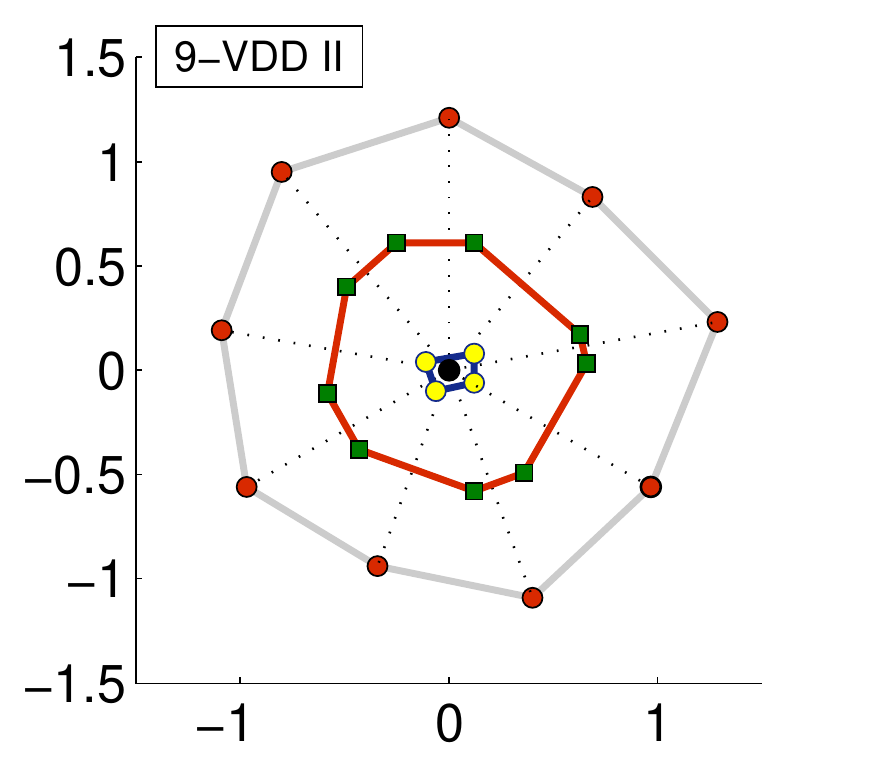}&
\includegraphics[width=.25\linewidth]{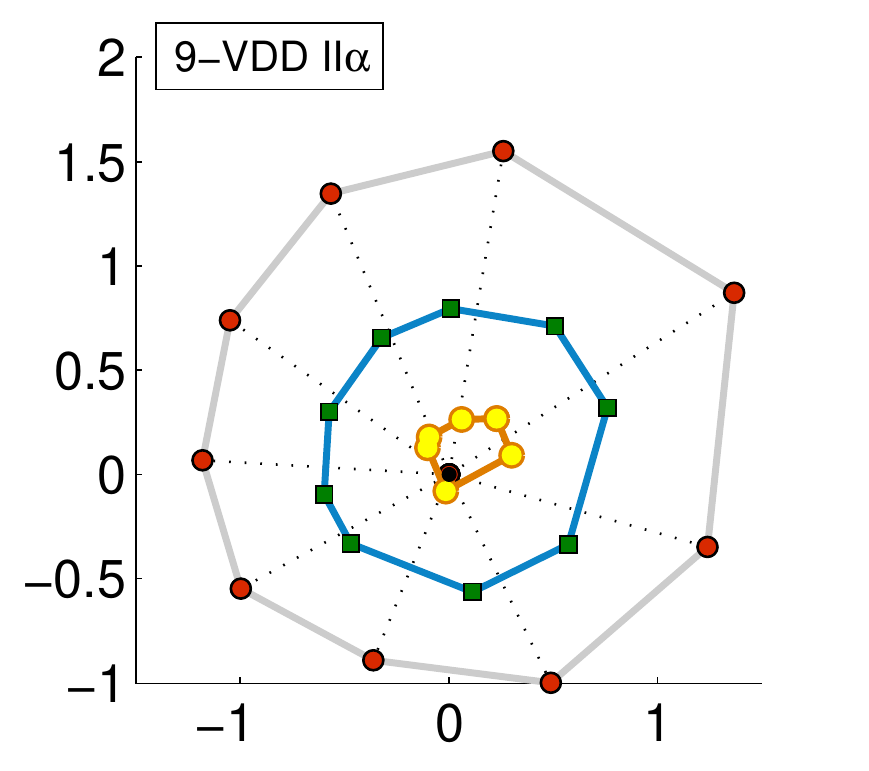}\\
\end{tabular}
\caption{Simulation examples of the $n$-VDD models II and II$_\alpha$ before scaling by ROI radius. \label{fig:model_II}}
\end{figure*}
}

{\setlength{\tabcolsep}{0pt}
\begin{figure*}[ht]
\centering
\footnotesize
\begin{tabular}{cccc}
\includegraphics[width=.25\linewidth]{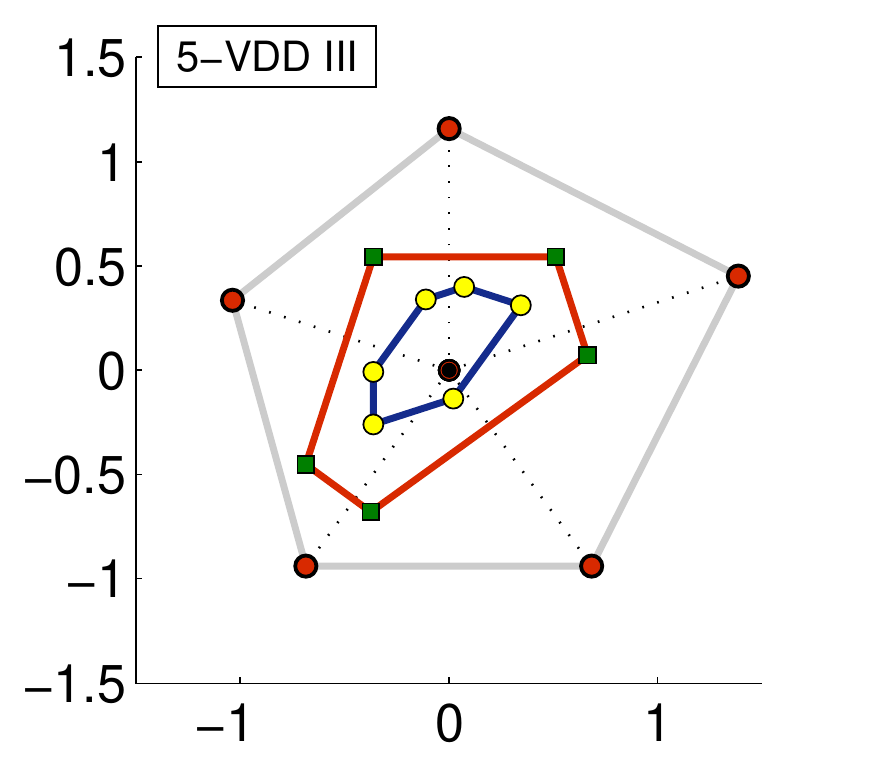}&
\includegraphics[width=.25\linewidth]{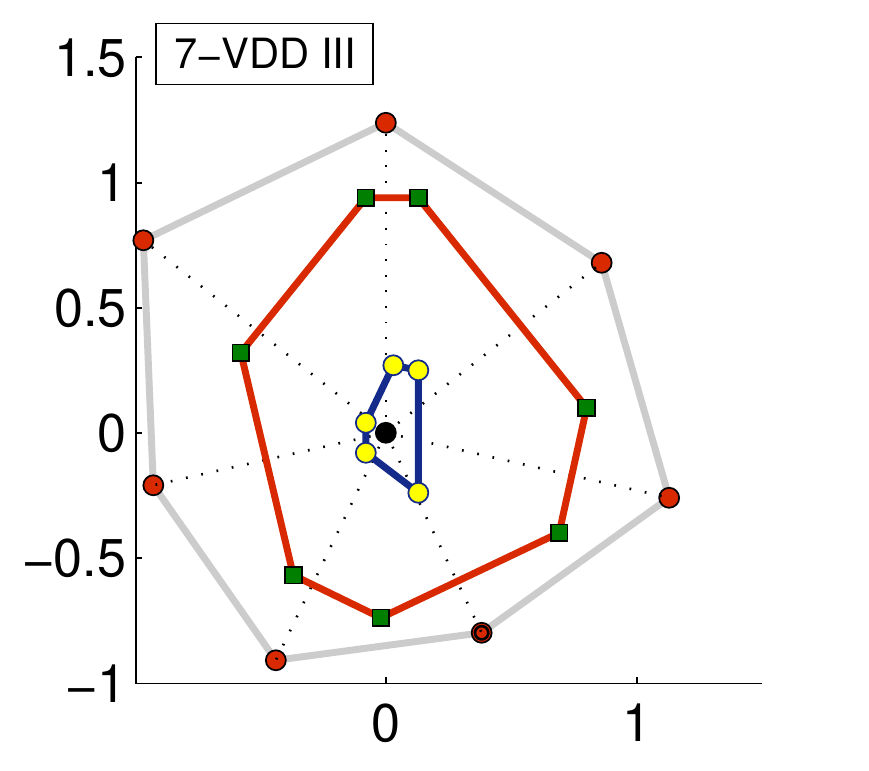}&
\includegraphics[width=.25\linewidth]{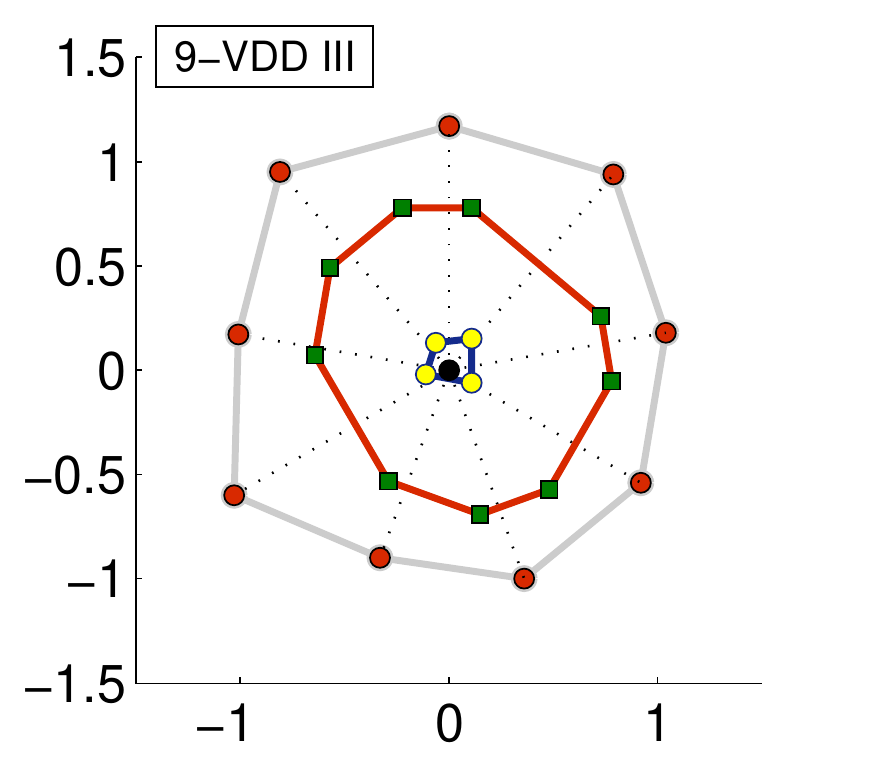}&
\includegraphics[width=.25\linewidth]{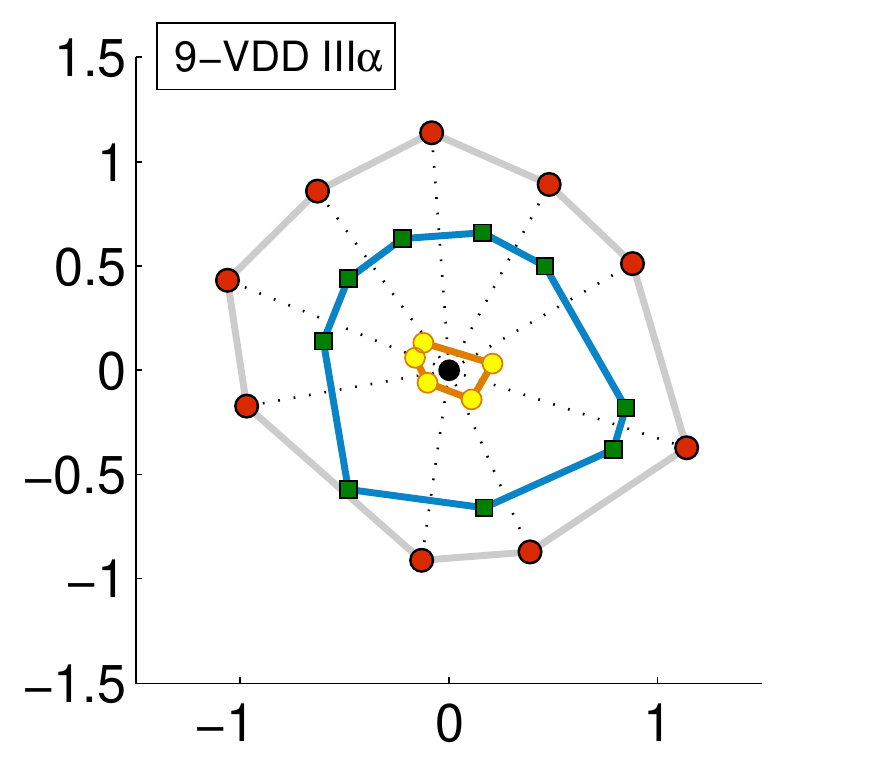}\\
\end{tabular}
\caption{Simulation examples of the $n$-VDD models III and III$_\alpha$ before scaling by ROI radius. \label{fig:model_III}}
\end{figure*}
}

In this framework, the scaled Voronoi cell $\mathbf{P_v^*}$ of the shifted Voronoi cell $\mathbf{P_v'}$ will be used as the \emph{concealing space} $\Omega_{cs}$ of the user, which hides the user location and is submitted to the LBS providers.
The area of the anonymity zone $\mathbf{A_z^*}$ resulted from the concealing space $\mathbf{P_v^*}$,  is defined as the \emph{privacy level}. Thus, bigger concealing space generates bigger anonymity zone and higher privacy level for the user location privacy.

{\setlength{\tabcolsep}{0pt}
\begin{figure}[ht]
\centering
\footnotesize
\begin{tabular}{c}
{\begin{overpic}[height=.5\linewidth]{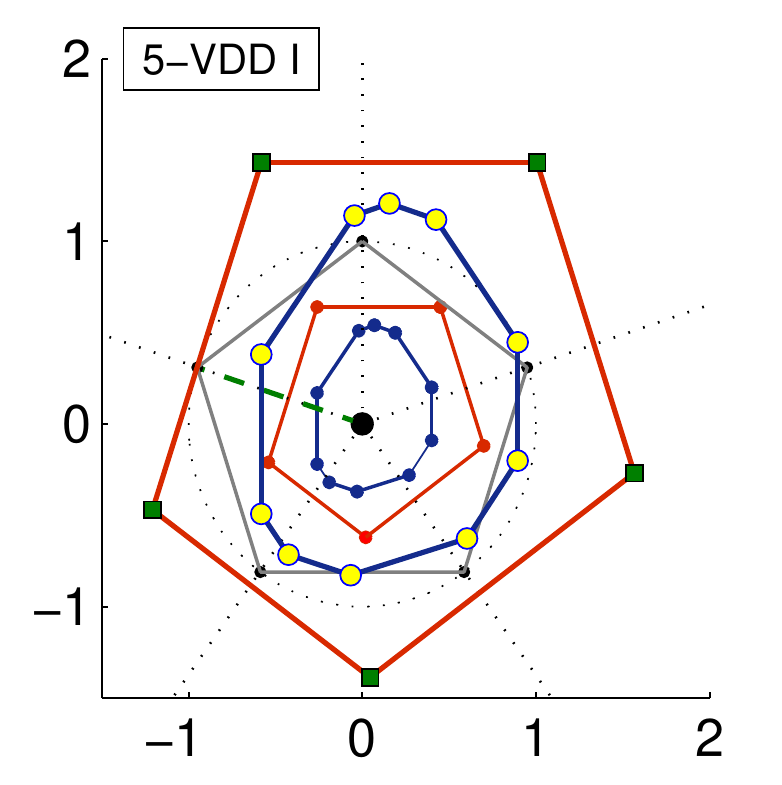}
\tiny
\cput(55,78){$\mathbf{P_v^*}$}
\cput(61,66){$\mathbf{A_z^*}$}
\cput(48,52){$\mathbf{A_z'}$}
\cput(29,45){$\mathbf{P_c}$}
\cput(60,52){$\mathbf{P_v'}$}
\cput(51,45){$O$}
\end{overpic}}
\\
\end{tabular}
\caption{Illustration of the VDD model I after scaling based on the ROI radius $r=1.0$. \label{fig:model_VDD}}
\end{figure}
}

When the user submits a query of places of interest within a circular range $r$, the query area is $\tau_0=\pi r^2$. We compute the closest distance $d_0$ to the edges of the originally generated $\mathbf{P_v'}$ and define the scalar $\lambda_v = \frac{r}{d_0}$, and then transform the Voronoi polygon $\mathbf{P_v'}$ by a scaling operation with scalar $\lambda_v$ and the user location $O$ as the origin. The transformed Voronoi polygon, i.e., the final concealing space, is denoted as  $\mathbf{P_v^*}=\lambda_v \mathbf{P_v'}$, which covers the query region with radius $r$, as shown in Fig. \ref{fig:model_VDD}.
Users also have flexibility to customize the privacy level in the system by setting the expected anonymity range $\iota$.
Then, similarly, we compute a scalar $\lambda_z = \frac{\iota}{d_0}$,
and transform the Voronoi polygon to be $\mathbf{P_v^*}=\lambda_z \mathbf{P_v'}$ by a $\lambda_z$-scaling around $O$.
If the user customizes both privacy level and concealing space, we use compute $\mathbf{P_v^*}=\max(\lambda_v, \lambda_z)\mathbf{P_v'}$.

All the computations are constructive by using planar geometry and linear algebra, and therefore are efficient.

\subsection{Contributions}
In this work, we present a novel location privacy framework, named $n$-VDD, based on the unique discrete geometric structure Voronoi-Delaunay Duality (VDD). In detail,
\begin{itemize}
\item Six models are proposed, to introduce irregularity to convex regular VDD structure.
\item The concealing space is derived from the Voronoi polygon. The privacy level can be customized.
\item The method is efficient by utilizing the planar VDD, where computations are linear line-line intersections.
\end{itemize}

The rest of the paper is organized as follows: Section \ref{sec:theory} describes the system model, Section \ref{sec:algorithm} presents the algorithms of the $n$-VDD models, Section \ref{sec:experiment} details the application settings, experiments and discussions, and finally Section \ref{sec:conclusion} concludes the paper.

\section{$n$-VDD Location Privacy Protection Model}
\label{sec:theory}
In this section, we first review the background knowledge of VDD, and then describe the VDD-based system for location privacy protection and the anonymizing protocol, and finally perform the attack analysis.

\subsection{Background}
\subsubsection{Voronoi-Delaunay Duality (VDD)}

Voronoi diagram and Delaunay triangulation are the classical geometric structures in computational geometry \cite{CGeomBook}. The Voronoi diagram of a point set $\mathbf{C}$ of $n$ points in the plane is a subdivision of the plane into $n$ Voronoi cells, such that each Voronoi cell around $C_i \in \mathbf{C}$ is the set of all points from which $C_i$ is the closest among all other points in $\mathbf{C}$. The dual of the Voronoi diagram is a unique triangulation, known as the Delaunay triangulation. A triangulation is \emph{Delaunay} means that it satisfies the empty circumcircle criterion, i.e., for any triangle, its circumcircle doesn't contain any other point, such triangle is called Delaunay triangle.
The optimal time complexity for constructing Voronoi diagram and Delaunay triangulation is $O(n \log n)$,  and the dual conversion between them costs $O(n)$.

\subsubsection{Local Structures} In our model, we require a proper exterior polygon such that the triangulation $T$ obtained by connecting each vertex of the polygon to the user location $O$ is Delaunay. Once the exterior polygon is fixed, the VDD structure is determined. Based on the VDD property, \emph{three convex polygons surrounding $O$} are generated, as follows:
\begin{itemize}
  \item \emph{Delaunay Polygon} $\mathbf{P_c}=\langle C_0C_1...C_{n-1}\rangle$, the convex boundary polygon of the Delaunay triangulation $\mathbf{T}$.
  \item \emph{Voronoi Polygon} $\mathbf{P_v}=\langle V_0V_1...V_{n-1}\rangle$, the convex polygon generated by applying the VDD principle on $\mathbf{T}$.

  \item \emph{Anonymity Zone} $\mathbf{A_z}=\langle A_0A_1...A_{n-1}\rangle$, the convex polygon generated by the intersection of the perpendicular strips for the edges of Voronoi polygon.
\end{itemize}

We start from the Voronoi-Delaunay structure of the convex regular $n$-sided polygon using $O$ as the center, and make variations to the structure to introduce irregularity. We adapt the exterior Delaunay polygon or/and interior Voronoi polygon to be irregular, which guarantees that the anonymity zone is a convex region, not a single point.
Different variations induce different anonymizing protocols.
The $n$-VDD framework may have other variations by introducing different irregularities.

\subsection{System Model}
\label{sub_sec:system_model}
In a typical LBS system, a user generates a query $q_0$ which is a tuple of his/her identification, the location $O=(x,y)$, the radius of the neighborhood $r$, and the points of interests (POI) $I$, such as the gas stations, ATMs and so on. That is,
$q_0 = \langle u_{id}, \langle(x,y), r \rangle, I \rangle$.
This query is transmitted to a local anonymizer engine which generates the concealing space (i.e., adapted Vononoi polygon), and based on the query framework, the concealing space is transmitted to LBS system.

\begin{itemize}
\item Model I: exterior polygon $\mathbf{P_c}$ is convex regular; interior Voronoi polygon is shifted to be irregular. The interior polygon generated by shifting is denoted as
\begin{equation}
\mathbf{P_v'}=\langle V_0'V_1'...V_{n-1}'\rangle.
\end{equation}

\item Model II: exterior convex regular polygon $\mathbf{P_c}$ is shifted to be irregular. The exterior polygon generated by shifting is denoted as
\begin{equation}
\mathbf{P_c'}=\langle C_0'C_1'...C_{n-1}'\rangle.
\end{equation}
The interior Voronoi polygon $\mathbf{P_v'}$ of $\mathbf{P_c'}$ is irregular.

\item Model III: exterior polygon is shifted to be irregular $\mathbf{P_c'}$; the interior Voronoi polygon of $\mathbf{P_c'}$ is shifted to $\mathbf{P_v'}$.

\end{itemize}

Suppose the area of the finally resulted interior polygon $\mathbf{P_v'}$ in each model is denoted as $\tau_v$.
According to the closest distance $d_0$ to the edges of the originally generated $\mathbf{P_v}$, we compute a scaling transformation of the $\mathbf{P_v'}$ with scalar $\lambda_v = \frac{r}{d_0}$ ($O$ as the origin), denoted as $\mathbf{P_v^*}=\lambda_v\mathbf{P_v'}$. $\mathbf{P_v^*}$ is the final concealing space,
\begin{equation}
\mathbf{P_v^*}=\langle V_0^*V_1^*\ldots V_{n-1}^*\rangle.
\end{equation}
Then the original query $q_0$ becomes
$
q_v=\langle u_{id}, \mathbf{P_v^*}, I\rangle.
$

The user may also set privacy level $\rho_0$, defined by the expected the anonymity radius parameter  $\iota$, $\rho_0=\pi \iota^2$.
In this case, the original query $q_0 = \langle u_{id}, \langle(x,y), r, \iota \rangle, I \rangle$.
Similarly, we compute a scalar $\lambda_z =
\frac{\iota}{d_0}$.
Then we have $\mathbf{P_v^*}=\max(\lambda_v, \lambda_z)\mathbf{P_v'}$.
If the user only cares about the customized privacy level, then $\mathbf{P_v^*}=\lambda_z\mathbf{P_v'}$.

\subsection{Anonymizing Protocol}
The anonymizing protocols are generally public to audience.
The followings are the common principles for all models:
\begin{itemize}
  \item \textsl{R0 - Scaling}: The submitted concealing space denoted as $\mathbf{P_v^*}$ is obtained by a $\lambda$-scaling of a convex polygon $\mathbf{P_v'}$ around the user location. The $\lambda$ is computed based on the user's requirements on privacy level and range of interest.

  \item \textsl{R1 - Voronoi-Delaunay Duality}: The convex polygon $\mathbf{P_v'}$
      has a dual convex polygon $\mathbf{P_c'}$: 1) each edge has a perpendicular dual edge; 2) all the dual edges intersect at the user location; and 3) all other endpoints of the dual edges are outside $\mathbf{P_v'}$ and form $\mathbf{P_c'}$, $\mathbf{P_c'}\supset\mathbf{P_v'}$.
    For each edge of $\mathbf{P_v'}$, a feasible strip is computed, which is perpendicular to the edge and exactly bounds it. The intersection of the feasible strips of all the edges is a convex \textit{feasible region}, denoted as $\mathbf{A_z}$. Similarly, the final anonymity zone $\mathbf{A_z^*}$ can be computed from $\mathbf{P_v^*}$. $\mathbf{A_z^*}$ differs from $\mathbf{A_z}$ by a scalar $\lambda$, which is $\lambda_z$ or $\lambda_v$.
\end{itemize}

The exterior polygon $\mathbf{P_c}$ can be convex regular (equiangular and equilateral) or irregular with different sector angles, then there are the following alternative principles:
\begin{itemize}
  \item \textsl{S0 - Sector Uniformization}:
  The polygon $\mathbf{P_c}$ is convex and regular taking the user location $O$ as the centroid, and forms a Delaunay triangulation by connecting each vertex of $\mathbf{P_c}$ to $O$, which defines sector rays. Then, the resulted dual Voronoi polygon $\mathbf{P_v}$ around $O$ is constructed by the perpendicular bisector intersections, and is regular.
  \item \textsl{S1 - Sector Shifting}: The sector rays of the original exterior polygon $\mathbf{P_c}$ are randomly shifted in a range to generate a new exterior polygon $\mathbf{P_c'}$, such that each vertex of the resulted dual Voronoi polygon $\mathbf{P_v'}$ is shifted within its corresponding sector.
\end{itemize}

Besides the above, each model has its own principle:

\begin{itemize}
\item \textsl{P1 - Interior Shifting}:
  The exterior polygon $\mathbf{P_c}$ and $O$ form a Delaunay triangulation by sector rays. The interior polygon $\mathbf{P_v'}$ is generated by randomly shifting each edge of the Voronoi cell $\mathbf{P_v}$ of $\mathbf{P_c}$ in parallel, such that each vertex of $\mathbf{P_v'}$ is shifted within its original sector.
  \item \textsl{P2 - Exterior Shifting}: The exterior polygon $\mathbf{P_c'}$ is generated by shifting each vertex of the original polygon $\mathbf{P_c}$ along its sector ray, such that each vertex of the resulted Voronoi polygon $\mathbf{P_v'}$ is shifted within its original sector.
  \item \textsl{P3 - Double Shifting} (\textsl{P1,P2}): The exterior polygon $\mathbf{P_c'}$ is generated by shifting the vertices of the original $\mathbf{P_c}$ along sector rays, and the resulted Voronoi edges are shifted in parallel to form the interior polygon $\mathbf{P_v'}$, such that each vertex of $\mathbf{P_v'}$ is shifted within its original sector. It is the combination of \textsl{P1} and \textsl{P2}.
\end{itemize}

Table \ref{tab:models} gives the anonymizing pipeline in each model.

\subsection{Attack Analysis}
\label{sub_sec:attack_analysis}
We analyze the ability of the VDD models to protect the user location $O$ from the attacker.
Suppose the query message $q_v$ is obtained by the attacker. Then the attacker has:
    \begin{itemize}
    \item An $n$-sided convex irregular polygon $\mathbf{P_v^*}$, which is a scaled (shifted) Voronoi polygon hiding the user location;
    \item An anonymizing protocol, one of Table \ref{tab:models}.
    \end{itemize}

\subsubsection{Protocol Attack}
Attacking can be tried by reversing the anonymizing process and analyzing each principle. The common principles are analyzed as follows:

\begin{itemize}
\item {\textsl{R0$^{-1}$ - Scaling:}}
The scaling uses the user location as the origin, which guarantees the user location is always within $\mathbf{P_v^*}$, and the shape of $\mathbf{P_v^*}$ is similar to that of the resulted interior polygon $\mathbf{P_v'}$ in all the models.

\item {\textsl{R1$^{-1}$ - Voronoi-Delaunay Duality:}}
For each edge of $\mathbf{P_v^*}$, we can find a feasible strip which is perpendicular to the edge and exactly bounds the edge. The intersection of all the feasible strips 
forms the \textit{anonymity zone} $\mathbf{A_z^*}$. For any point $o\in \mathbf{A_z^*}$, we find the point $C_i^*$ for edge $V_{i-1}^*V_i^*$ of $\mathbf{P_v^*}$ such that $V_{i-1}^*V_i^*$ is the perpendicular bisector of $oC_i^*$. All the $C_i^*$'s form the exterior Delaunay polygon $\mathbf{P_c^*}$ (see Fig. \ref{fig:VDD}). The Delaunay edges $C_{i-1}^*C_i^*$ exist and are uniquely determined. Therefore, every point in the anonymity zone has a VDD structure satisfying principle \textsl{R1}, and could be the user location $O$.

\end{itemize}

Principles specific to each model are analyzed as follows:

\begin{itemize}
\item {\textsl{S0$^{-1}$ - Sector Uniformization}:}
From \textsl{S0}, the surrounding angles around the user location are identical to be $\frac{2\pi}{n}$, where $n$ is the number of vertices of $\mathbf{P_v^*}$ (convex).
For any point $o\in \mathbf{A_z^*}$, we draw the perpendicular lines to all the edges of $\mathbf{P_v^*}$ and the angles surrounding $o$ won't change when we shift the point within $\mathbf{A_z^*}$.

\item {\textsl{(S0, P1)$^{-1}$ - Interior Shifting}:}
From \textsl{P1}, irregular $\mathbf{P_v^*}$ is obtained by shifting the regular Voronoi polygon, which takes the user location as the centroid/center. Given any point $o\in \mathbf{A_z^*}$, by shifting back the edges of $\mathbf{P_v^*}$, it is guaranteed to generate a regular polygon to induce a Delaunay polygon by \textsl{R1}. Therefore, every point in the anonymity zone satisfies principles \textsl{(S0, P1)} (Model I).

\item {\textsl{(S0, P2)$^{-1}$ - Exterior Shifting}:}
For any point $o\in \mathbf{A_z^*}$, the Delaunay polygon $\mathbf{P_c^*}$ resulted from the irregular $\mathbf{P_v^*}$ is unique and irregular (by \textsl{R1}). From \textsl{S0}, all surrounding angles at $o$ are equal. Then by shifting the edges of $\mathbf{P_c^*}$, it is guaranteed to generate a regular polygon taking $o$ as the centroid. That means every point in the anonymity zone satisfies principles \textsl{(S0, P2)} (Model II).

\item {\textsl{(S0, P3)$^{-1}$ - Double Shifting}:}
For any point $o\in \mathbf{A_z^*}$, the Delaunay polygon $\mathbf{P_c^*}$ is uniquely computed (by \textsl{R1}).
We shift $\mathbf{P_v^*}$ with a random range to $\mathbf{P_v'}$ and then update the Delaunay polygon to be $\mathbf{P_c'}$. By shifting $\mathbf{P_c'}$, it is guaranteed to generate a regular polygon which takes $o$ as the centroid. That means every point in the anonymity zone satisfies principles \textsl{(S0, P3)} (Model III).
\end{itemize}

\begin{itemize}
\item {\textsl{S1$^{-1}$ - Sector Shifting}:}
From \textsl{S1}, the surrounding angles around the user location are unequal, and $\mathbf{P_v^*}$ is irregular.
For any point $o\in \mathbf{A_z^*}$, we compute the Delaunay polygon $\mathbf{P_c^*}$ of $\mathbf{P_v^*}$ by \textsl{R1}. The angles around $o$ can be computed and won't change if shifting the point within $\mathbf{A_z^*}$.

\item {\textsl{(S1, P1)$^{-1}$ - Sector \& Interior Shifting}, \textsl{(S1, P2)$^{-1}$ - Sector \& Exterior Shifting}, \textsl{(S1, P3)$^{-1}$ - Sector \& Double Shifting}:}
For any point $o\in \mathbf{A_z^*}$, the Delaunay polygon $\mathbf{P_c^*}$ is uniquely computed (by \textsl{R1}).
Shifting $\mathbf{P_v^*}$, $\mathbf{P_c^*}$, or both won't influence the surrounding angles, and also won't generate a regular polygon. Therefore, every point in the anonymity zone satisfies corresponding principles (Models I$_\alpha$-III$_\alpha$).
\end{itemize}

Therefore, in all Models I-III, I$_\alpha$-III$_\alpha$, every point in the anonymity zone could be the user location, and
the attackers cannot differentiate the points in the anonymity zone.

\subsubsection{Centroid Attack}
Centroid is easy to compute for a given polygon. Here, we need analyze whether the centroid can be used for the attack.

{\setlength{\tabcolsep}{0pt}
\begin{figure}[t]
\centering
\footnotesize
\begin{tabular}{cc}
\includegraphics[height=.45\linewidth]{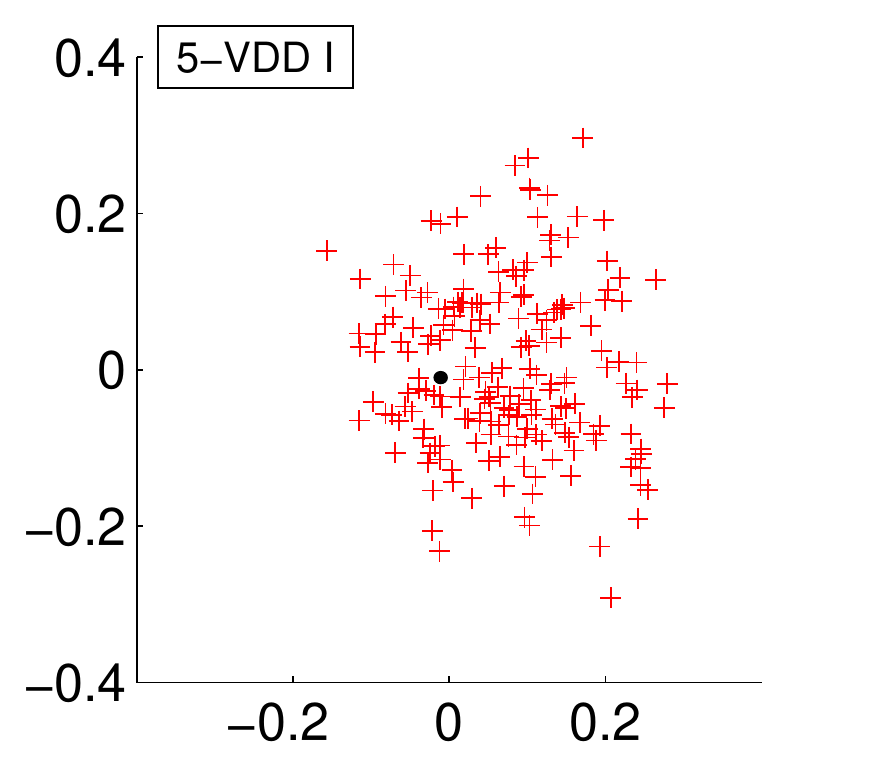}&
\includegraphics[height=.45\linewidth]{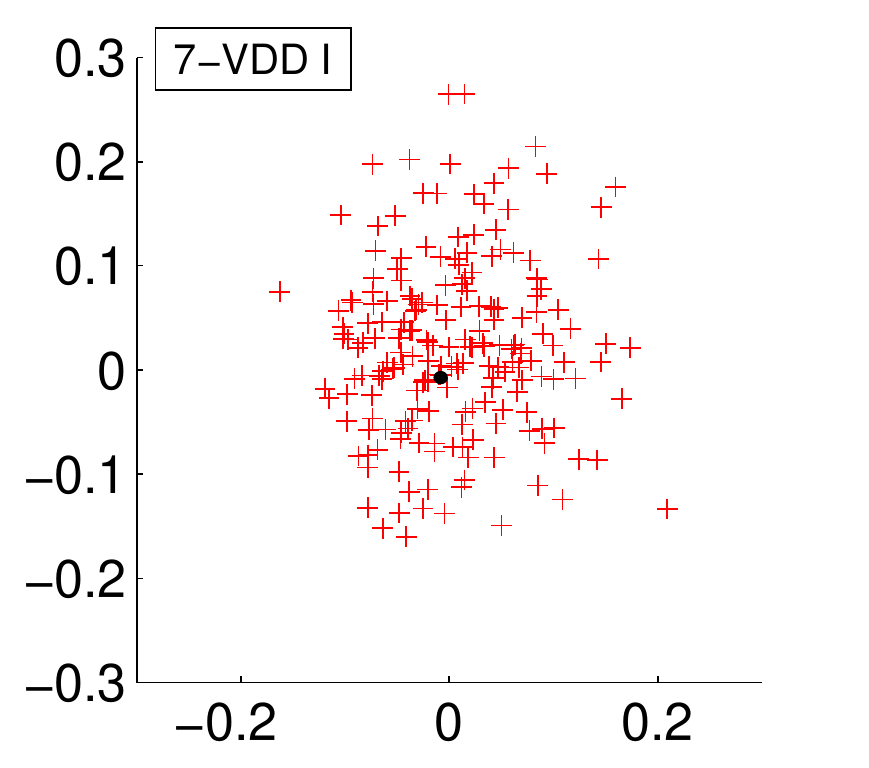}\\ \includegraphics[height=.45\linewidth]{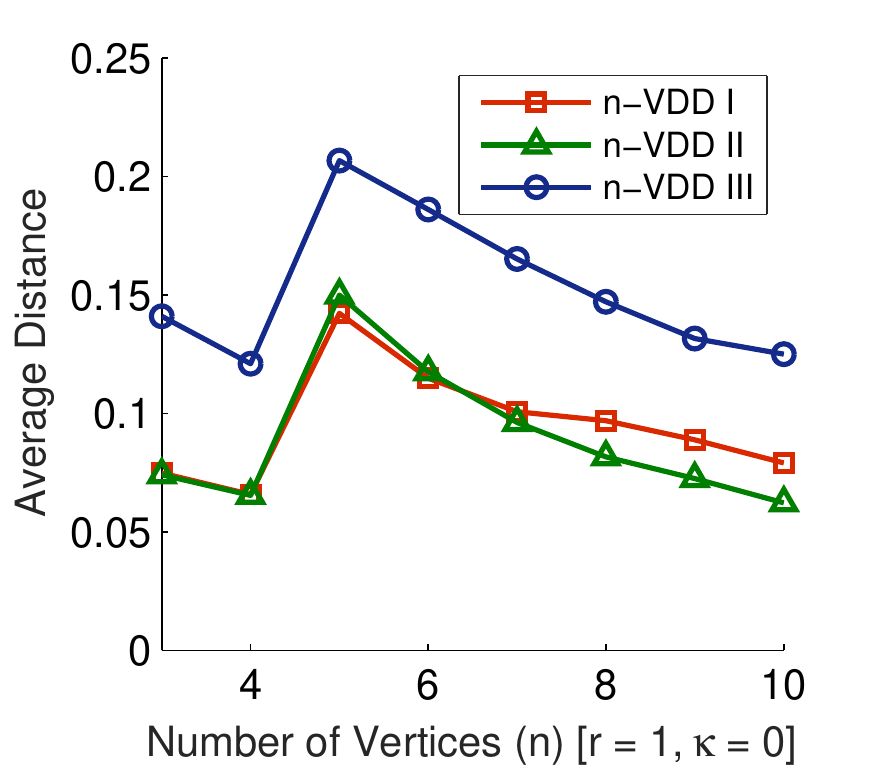}&
\includegraphics[height=.45\linewidth]{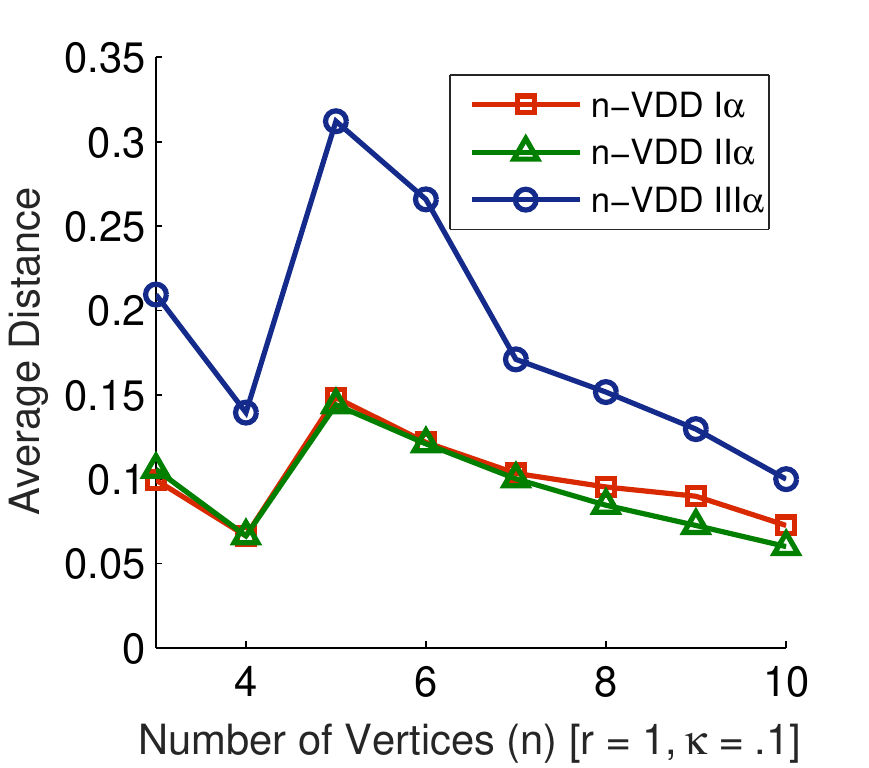}
\end{tabular}
\caption{Distributions of centroids of Voronoi polygons (see two examples in top row) and the plots of the average distance from the seed $O$ (bottom row).}
\label{fig:vdd_controids}
\end{figure}
}

First of all, all the anonymizing protocols introduce the irregularity, so that the centroid of the submitted Voronoi polygon $\mathbf{P_v^*}$ (concealing space), denoted as $C_{Pv}$, is not guaranteed to coincide with the user location. In Models I-III, $\mathbf{P_v^*}$ is with equal sector angles, and the centroid of interior/exterior regular polygon in a VDD structure gives the user location, however, the centroid of irregular $\mathbf{P_v^*}$ is not the user location. In Models I$_\alpha$-III$_\alpha$, there is no guarantee that the centroid of irregular $\mathbf{P_v^*}$ is the user location. Therefore, direct concealing space centroid attack can be avoided for each model. To verify this, we generated $1000$ irregular polygons by random shifting the vertices along the sector rays (exterior vertex shifting), and computed the centroids. Figure \ref{fig:vdd_controids} gives two examples of the distribution of the centroids around the user location $O$ (top), and also plots the average distance from $O$ for $n=3..10$. It is observed that the centroid $C_{Pv}$ is away from the user location $O$ in general case.

{\setlength{\tabcolsep}{0pt}
\begin{figure}[h]
\centering
\begin{tabular}{cc}
\includegraphics[height=.4\linewidth]{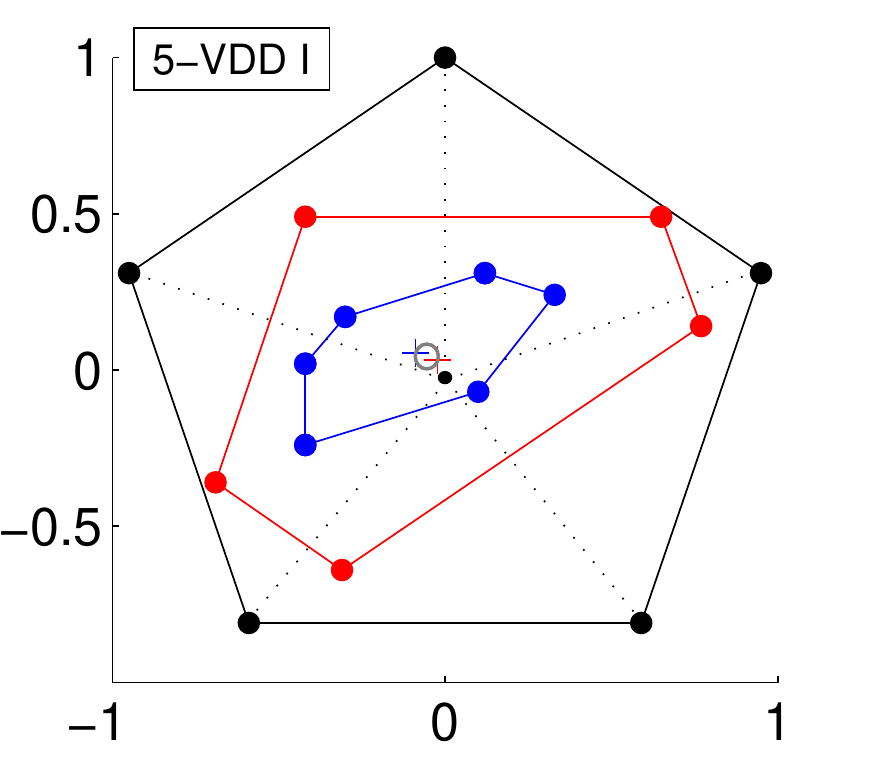}&
\includegraphics[height=.4\linewidth]{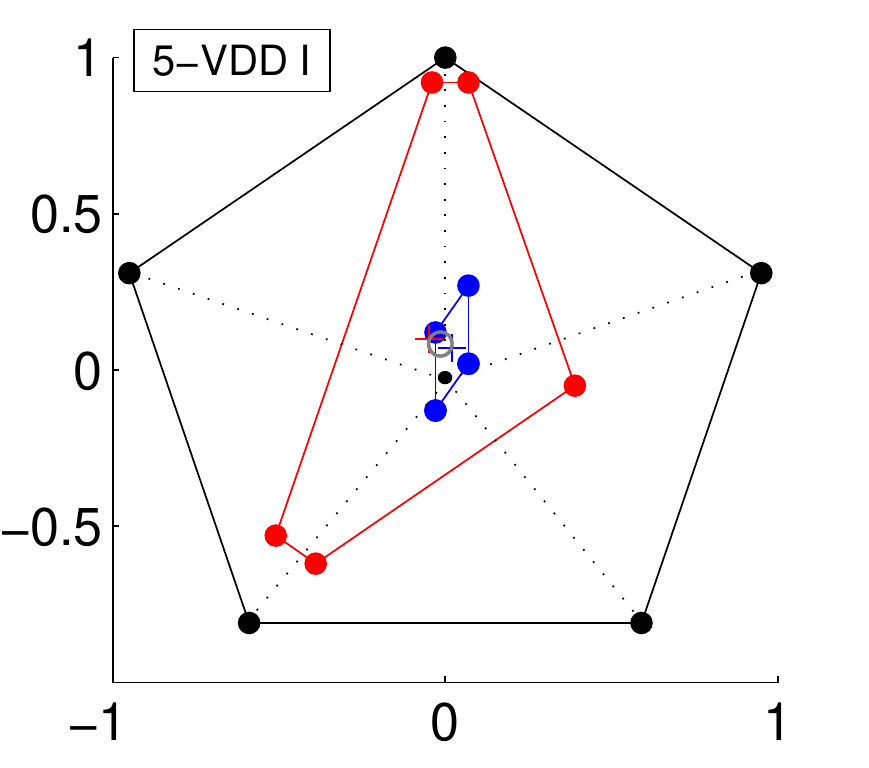}
\end{tabular}\\
\includegraphics[height=.4\linewidth]{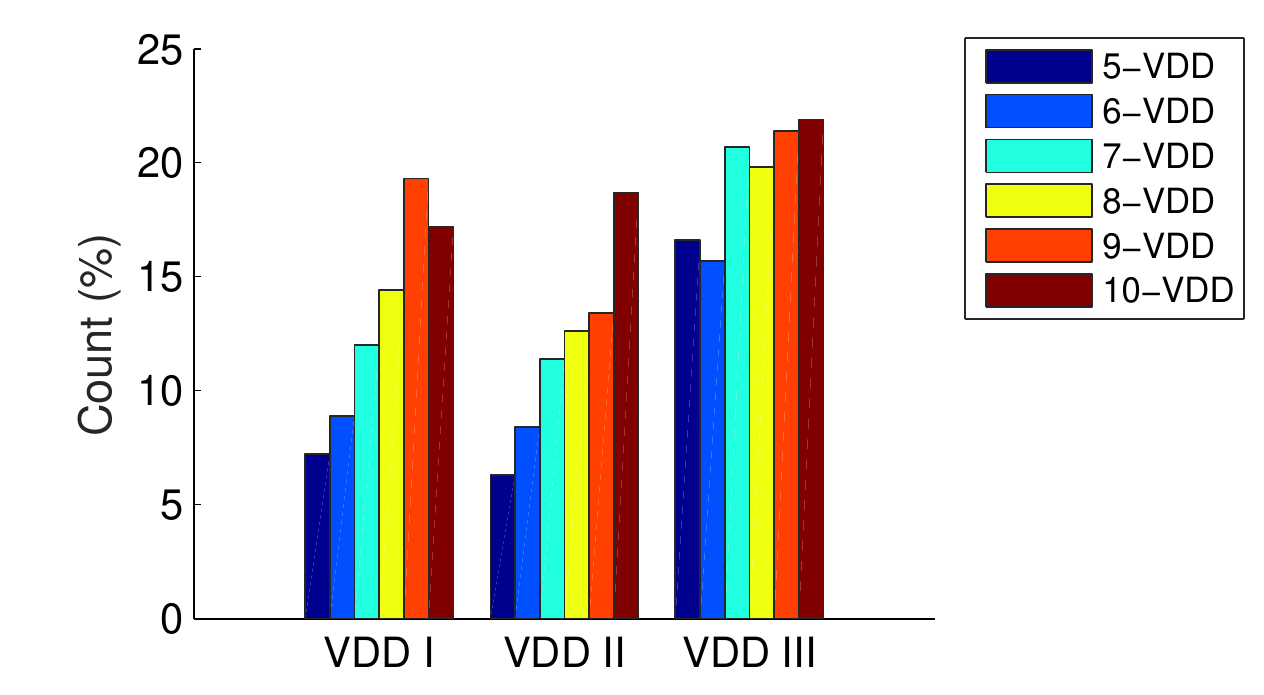}\\
\caption{The percentage (\%) of times the seed $O$ falls in the circle of using the centroids of concealing space (red `+') and anonymity zone (blue `+') as the diameter (see top row).
\label{fig:vdd_centroids_AZ}}
\end{figure}
}

After generating the anonymity zone $\mathbf{A_z^*}$ from $\mathbf{P_v^*}$, is it possible to use the centroid of $\mathbf{A_z^*}$, denoted as $C_{Az}$, to reveal the user location? Similarly, $C_{Az} \in \mathbf{A_z^*}$, $C_{Az}$ is not guaranteed to coincide with the user location. Similar experiments verified that. Then, will the two centroids, $C_{Pv}$ and $C_{Az}$, give a hint to shrink the range? First, $C_{Pv}$ can be inside or outside $C_{Pv}$, there is no fixed relationship between their positions, as shown in Fig. \ref{fig:vdd_centroids_AZ} (top). We then create a circle passing through the two centroids and using the segment between them as the diameter, and detect whether the circle includes the user location $O$. Figure \ref{fig:vdd_centroids_AZ} (bottom) shows that the only a low portion (less than $23\%$) of the 1000 examples in the cases of $n=5..10$ has the user location in the circle of the two centroids. That means the circle cannot be used to replace or shrink $\mathbf{A_z^*}$.

\section{Computational Algorithms}
\label{sec:algorithm}

This section details the computation of the anonymizing process and the anonymity zone for the proposed models.
They share the same algorithms for the common anonymizing principles \textsl{R0} (scaling) , \textsl{R1} (VDD), and have different algorithms for their specific principles.

\subsection{Common Algorithms}

\subsubsection{Voronoi Polygon}
The Voronoi polygon $\mathbf{P_v}$ is generated from the Delaunay triangulation $\mathbf{T}$ formed by the Delaunay polygon $\mathbf{P_c}$ and the seed $O$ (user location). Each vertex $V_i$ of $\mathbf{P_v}$ is the intersection of perpendicular bisector lines of line segments $\overline{OC_{i}}$ and $\overline{OC_{i+1}}$ ($\overline{OC_{n}}=\overline{OC_{0}}$) (see Fig. \ref{fig:VDD}).

\subsubsection{Anonymity Zone}
From a given Voronoi polygon $\mathbf{P_v}$, the anonymity zone (feasible Region) $\mathbf{A_z}$ is generated by computing the intersecting region of the parallel strips, which are perpendicular to and bounded by the Voronoi edge $\overline{V_iV_{i+1}}$. Any point in $\mathbf{A_z}$ could be a feasible solution to the seed $O$. In detail, assume $\overline{L_i}$, $\overline{L_{i+1}}$ are the perpendicular lines at $V_i$, $V_{i+1}$, respectively, and the parallel strip is denoted as $\mathbf{\Gamma}_{i+1}= \overline{L_i} \wedge  \overline{L_{i+1}}$. Then $\mathbf{A_z}=\cap \{\mathbf{\Gamma}_{i}\}, \forall i=0, \ldots,n-1$.

\subsubsection{Scaling}

After getting the interior (shifted) Voronoi polygon $\mathbf{P_v'}$ and its anonymity zone $\mathbf{A_z'}$, we perform a scaling transformation using the user location $O$ as the origin with a scaling factor $\lambda$. $\lambda$ can be computed in different ways according to the user query about the ROI radius or the privacy level or both, as reported in Section \ref{sub_sec:system_model}.
Therefore, the obtained concealing space $\mathbf{P_v^*}=\lambda \mathbf{P_v'}$.

\begin{algorithm}[t]
\small
\caption{Model I - Interior Shifting \label{alg:model_1}}
\begin{algorithmic}
\REQUIRE User location $O$, vertex number $n$, radius of interest $r$
\ENSURE The concealing space $\mathbf{P_v^*}$
\STATE Compute a regular $n$-sided polygon $\mathbf{P_c}=\langle C_0C_1\ldots C_{n-1}\rangle$, such that the centroid $\frac{\Sigma C_i}{n}=O$
\STATE Construct the triangulation $\mathbf{T}$ by connecting each $C_i$ to $O$
\STATE Compute the Voronoi cell $\mathbf{P_v}=\langle V_0V_1\ldots V_{n-1}\rangle$ of $\mathbf{T}$ 
\STATE $e_0' \leftarrow \overline{V_0V_1}$
\FOR{$i \leftarrow 1$ \textbf{\textit{to}} $n-1$}
\STATE Compute the feasible shifting range $\tau_i$ for 
$e_{i+1}=V_iV_{i+1}$ 
\STATE Shift $e_{i+1}$ in parallel to $Rand(\tau_i)$ and get edge $e_{i+1}'$
\STATE $V_i' \leftarrow e_{i+1}' \cap e_{i}'$
\ENDFOR
\STATE Compute the feasible region of 
 $\mathbf{P_v'}=\langle V_0'V_1'\ldots V_{n-1}'\rangle$
\STATE Compute the closest distance to the edges $e_i'$ and compute the scaling by scalar
 $\lambda = \frac{r}{d_0}$
\STATE Return $\mathbf{P_v^*}\leftarrow \lambda \mathbf{P_v'}$
\end{algorithmic}
\end{algorithm}

\subsection{Model I - Interior Shifting}
 Model I introduces the irregularity to the interior Voronoi polygon by shifting the Voronoi edges in their corresponding feasible ranges to guarantee the Voronoi vertices are still in their original sectors (to grantee $O\in \mathbf{A_z'}$).

 Algorithm \ref{alg:model_1} shows the computation pipeline.
 We first generate a regular polygon using the user location as the centroid and compute the Voronoi polygon based on Voronoi-Delaunay duality, then shift the Voronoi edges to generate an irregular Voronoi polygon, and finally scale the shifted Voronoi polygon and submit the result to LBS providers.

In detail, after getting the regular Voronoi polygon, we fix the first Voronoi edge $e_0'=e_0$, and shift the left Voronoi edges $e_i=\overline{V_iV_{i+1}}$ to the new one $e_i'$ one by one. For each Voronoi edge $e_i$, the shift range is determined by the previous Voronoi edge $e_{i-1}'$. For the last Voronoi edge $e_{n-1}$, the shift range is determined by both $e_0$ and $e_{n-2}'$.

\begin{figure}[h]
\includegraphics[width=0.35\linewidth]{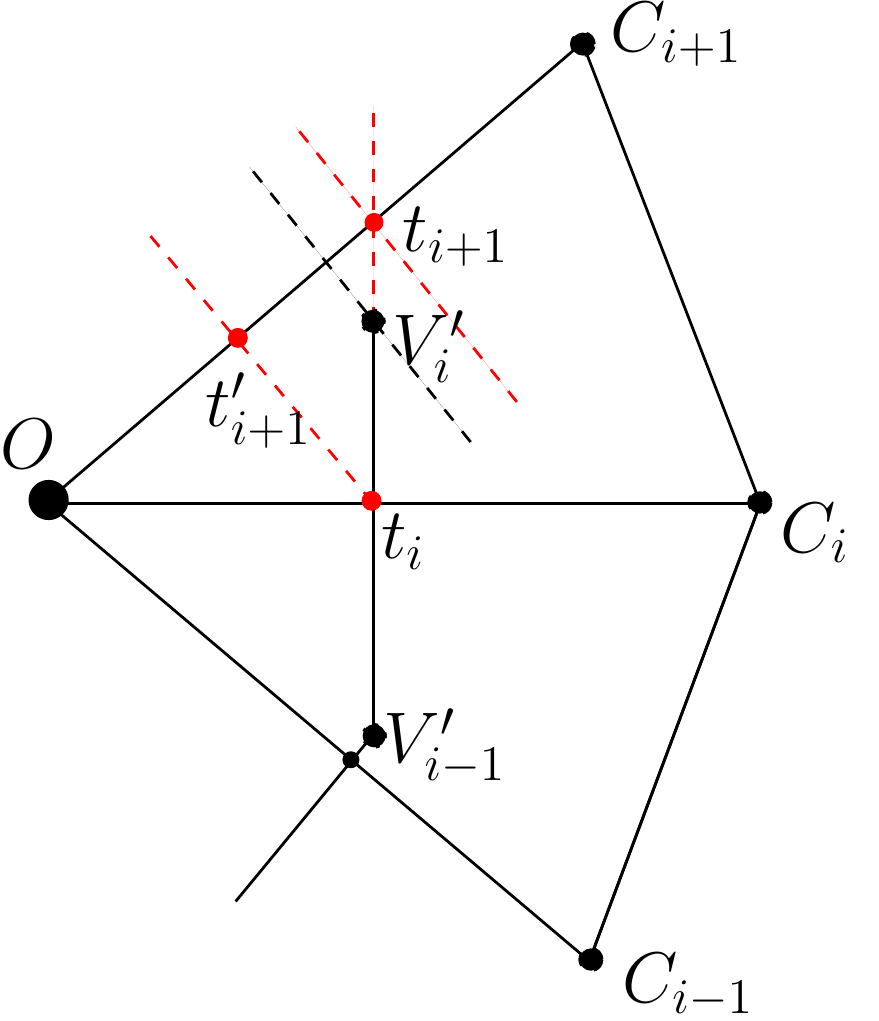}
\includegraphics[width=0.35\linewidth]{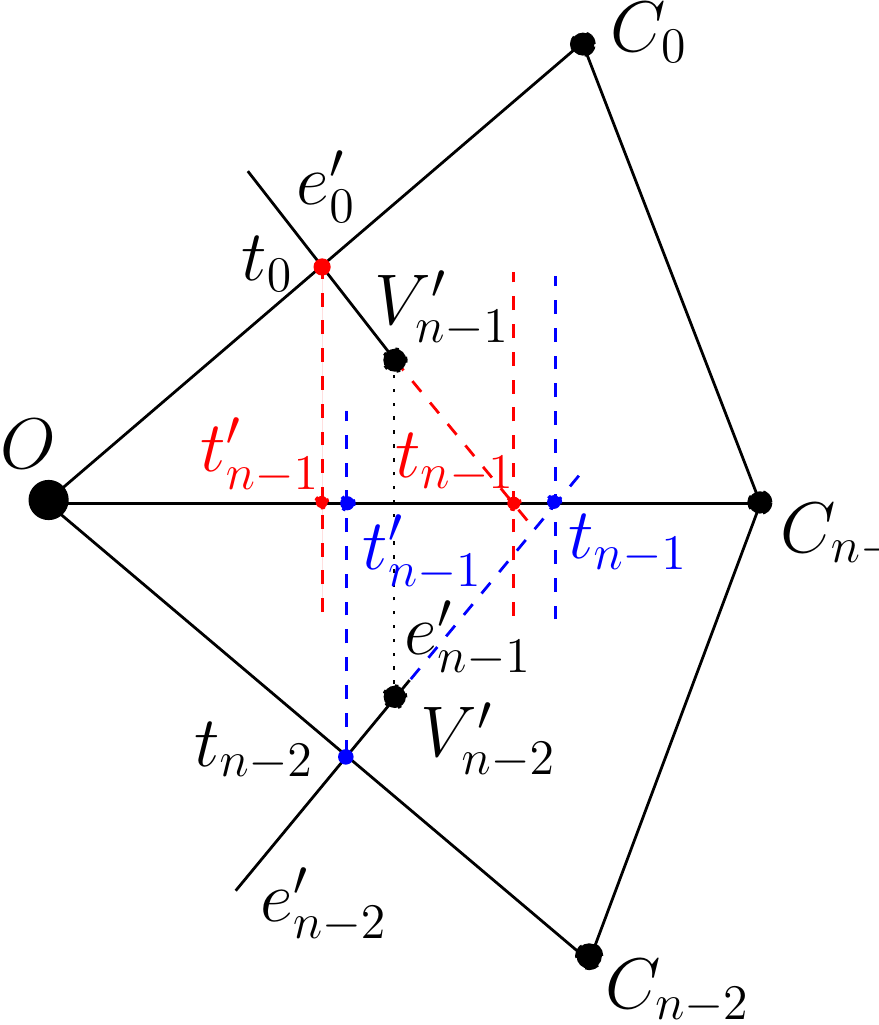}
\centering
\caption{Range for Voronoi edge shifting.}
\label{fig:range_1}
\end{figure}

As shown in Fig. \ref{fig:range_1}, $e_i'$ is extended to intersect $\overline{OC_i}$ and $\overline{OC_{i+1}}$ at $t_i$, $t_{i+1}$, respectively. From $t_i$, we issue a ray perpendicular to $\overline{OC_{i}}$ which intersects $\overline{OC_{i+1}}$ at $t_{i+1}'$. Then the range $\tau_{i+1}=[t_{i+1}', t_{i+1}]$. The position witihn this range can guarantee the new Voronoic polygon vertex $V_i'=e_i'\cap e_{i+1}'$ is within the Delaunay triangle $\bigtriangleup C_iOC_{i+1}$. For the last Voronoi edge $e_{n-1}$, we compute the range $\tau_{n-1}$ determined by $e_{n-2}'$, and the range $\tau_0$ determined by $e_0'$. Therefore, $\tau_{n-1}\leftarrow \tau_{n-1} \cap \tau_0$. The uniformly and randomly selected Voronoi edge $e_{n-1}'$ within the range $\tau_{n-1}$ intersects $e_{n-2}'$ and $e_0'$ and produces $V_{n-2}'$ and $V_{n-1}'$, respectively. Then the resulted shifted Voronoi cell is $\mathbf{P_v'}=\langle V_0'V_1'\ldots V_{n-1}'\rangle$.

For the case of $n=3$ or $4$, the Voronoi polygon is a triangle or rectangle, then the anonymity zone is itself. They give very strong and interesting results, where the attacker can do nothing for predicting the user location. In addition, the probability to generate a regular shifted Voronoi polygon with $n$ random numbers is very low and almost won't happen. We never met this situation in our large amounts of experiments. In order to make it for sure, one may add a regularity test to avoid this case. 
Figure \ref{fig:model_I} shows the simulation results, which include three layers of polygons, $\mathbf{P_c}$, $\mathbf{P_v'}$ and $\mathbf{A_z'}$.
Figure \ref{fig:interior_shifting} gives the histograms of the distances from the shifted Voronoi edge to the user location $O$ by generating 200 5-sided polygons, to show the irregularity generated by Voronoi edge (interior) shifting. Note that in the original convex regular polygon $\mathbf{P_c}$ with sector radius 1.0, the distance of Voronoi edge to the center $O$ is identical to be 0.5.

{\setlength{\tabcolsep}{0pt}
\begin{figure}[h]
\centering
\footnotesize
\begin{tabular}{cc}
\includegraphics[height=.45\linewidth]{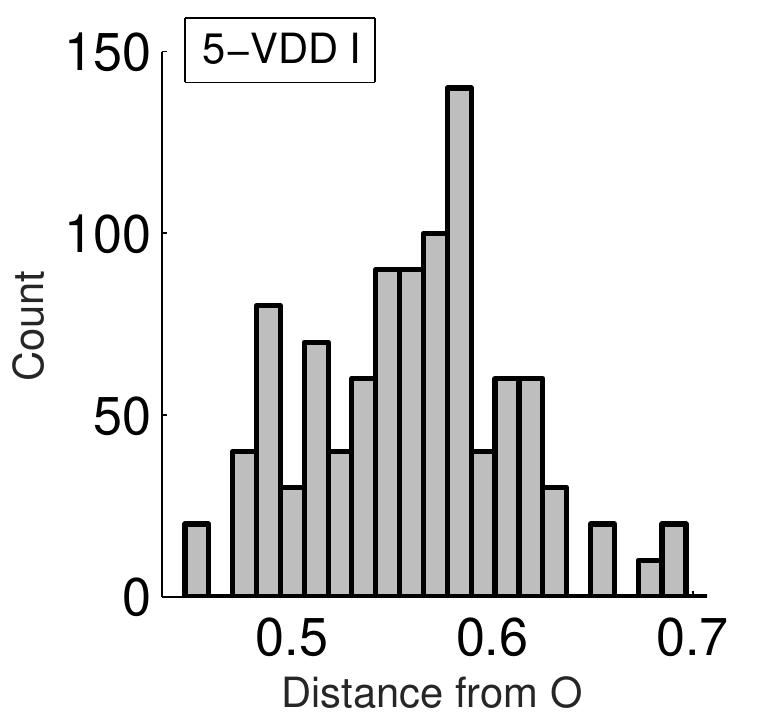}&
\includegraphics[height=.45\linewidth]{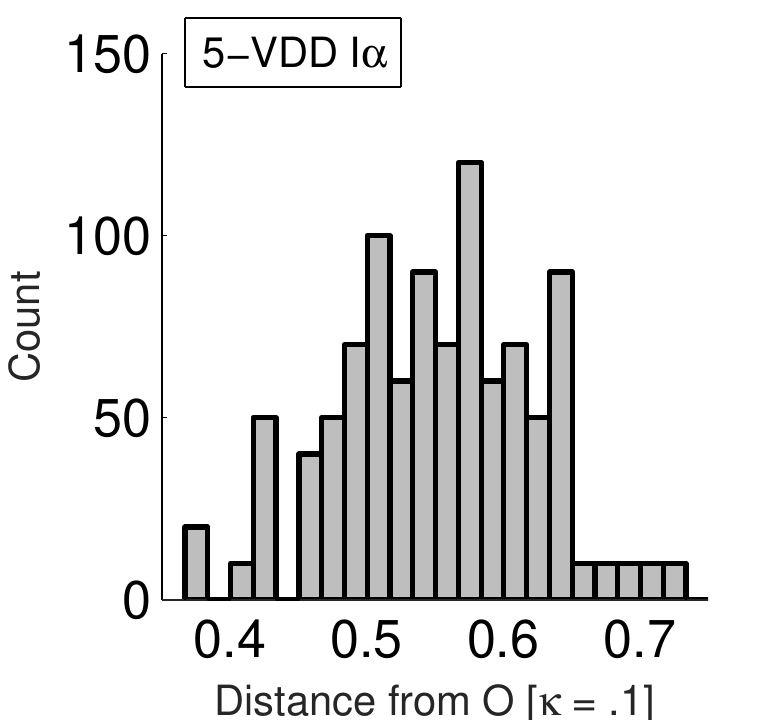} \\
\end{tabular}
\caption{Distances of Voronoi edge shifted from $O$.\label{fig:interior_shifting}}
\end{figure}
}

\subsection{Model II - Exterior Shifting}
Model II introduces irregularity to the exterior Delaunay polygon by shifting the regular Delaunay polygon vertices along their corresponding sector rays within a shifting feasible range, so that the resulted Voronoi polygon is irregular.

Algorithm \ref{alg:model_2} gives the computation pipeline. We first generate an irregular Delaunay polygon around the user location, then compute the Voronoi polygon, 
and finally scale the Voronoi polygon and submit the result to LBS provides.

We first select one point in the neighborhood of the user location $O$ as the vertex $C_0'$. We generate the vertices $C_i'$ one by one. Here, we set the sector angles to be equal, $\alpha_i=\alpha=\frac{2\pi}{n}$. The range for $C_{i+1}'$ is determined by $C_i'$ to guarantee the Voronoi vertex $V_i'$ is within the triangle $\bigtriangleup C_i'OC_{i+1}'$. Similarly, the last vertex $C_{n-1}'$ is determined by both $C_0'$ and $C_{n-2}'$.

\begin{algorithm}[t]
\small
\caption{Model II - Exterior Shifting \label{alg:model_2}}
\begin{algorithmic}
\REQUIRE User location $O$, vertex number $n$, radius of interest $r$
\ENSURE The concealing space $\mathbf{P_v^*}$
\STATE Select a point $C_0$ around $O$ to form the 0-th sector ray
\STATE set sector angles $\alpha_i=2\pi/n$
\FOR{$i \leftarrow 1$ \textbf{\textit{to}} $n-1$}
\STATE Compute the feasible shifting range $\tau_i$ for $C_i$
\STATE $C_i' \leftarrow Rand(\tau_i)$

\ENDFOR
\STATE Construct the irregular $n$-sided polygon $\mathbf{P_c'}=\langle C_0'C_1'\ldots C_{n-1}'\rangle$
\STATE Construct the triangulation $\mathbf{T}$ by connecting each $C_i$' to $O$
\STATE Compute the Voronoi cell $\mathbf{P_v'}=\langle V_0'V_1'\ldots V_{n-1}'\rangle$ of $\mathbf{T}$ \STATE Compute the closest distance to the edges $e_i'$ and compute the scaling by scalar
 $\lambda = \frac{r}{d_0}$
\STATE Return $\mathbf{P_v^*}\leftarrow \lambda \mathbf{P_v'}$
\end{algorithmic}
\end{algorithm}

\begin{figure}[h]
\includegraphics[width=0.4\linewidth]{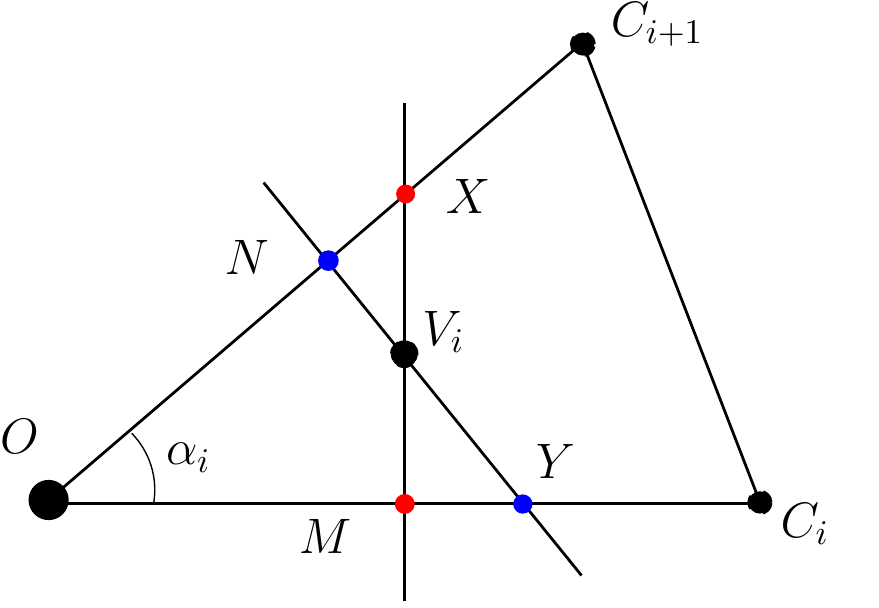}
\centering
\caption{Range for Delaunay polygon vertex selection.}
\label{fig:range_2}
\end{figure}

As shown in Fig. \ref{fig:range_2}, by VDD, we have
$ |\overline{OM}| = \frac{1}{2}|\overline{OC_{i}'}|$,
$ |\overline{ON}| = \frac{1}{2}|\overline{OC_{i+1}'}|   $,
$|\overline{XM}|\perp |\overline{OC_{i}'}| $, and
$|\overline{YN}|\perp |\overline{OC_{i+1}'}| $. In addition, $|\overline{OM}|=|\overline{OX}|\cos{\alpha_i}$, and
$|\overline{ON}|=|\overline{OY}|\cos{\alpha_i}$.
In order to keep the intersection of $|\overline{OC_{i}'}|$ and $|\overline{OC_{i+1}'}|$ in the sector $\sphericalangle  C_{i+1}'OC_{i}' $, the following conditions must hold:

\begin{enumerate}
  \item $ |\overline{ON}|<|\overline{OX}|$: then   $|\overline{ON}|<\frac{|\overline{OM}|}{\cos{\alpha_i}}$
$\Rightarrow |\overline{ON}|<\frac{|\overline{OC_{i}}|}{2cos{\alpha_i}}$
$\Rightarrow \frac{|\overline{OC_{i+1}}|}{2}<\frac{|\overline{OC_{i}}|}{2\cos{\alpha_i}}$
$\Rightarrow |\overline{OC_{i+1}}|<\frac{|\overline{OC_{i}}|}{\cos{\alpha_i}}$.

  \item $|\overline{OM}|<|\overline{OY}|$: then  $|\overline{OC_{i+1}'}|>|\overline{OC_{i}'}\cos{\alpha_i}|$,  similarly.
\end{enumerate}

So, we have the inequality
\begin{equation}
|\overline{OC_{i}'}| \cos{\alpha_i} < |\overline{OC_{i+1}'}| < \frac{|\overline{OC_{i}'}|}{\cos{\alpha_i}}. \end{equation}

For $n\geq5$, $\alpha_i<\pi/2$. For $n=3, 4$, the angle is close to or greater than $\pi/2$, then the upper bound will be very large or no intersection. So we give a parameter $\mu$ to constrain the range. Therefore, we have
\begin{equation}
|\overline{OC_{i}'}| \cos{\alpha_i} < |\overline{OC_{i+1}'}| < \min(\mu |\overline{OC_{i}'}|, \frac{|\overline{OC_{i}'}|}{\cos{\alpha_i}}).
\end{equation}
For the last vertex $C_{n_1}'$, we take the intersection of the ranges based on $C_0',C_{n-2}'$ using the similar strategy. Then the resulted irregular Delaunay polygon is $\mathbf{P_c'}=\langle C_0'C_1'\ldots C_{n-1}'\rangle$. Similarly, for $n=3,4$, $\mathbf{P_c'}$ is a triangle or rectangle (see Fig. \ref{fig:VDD}), and $\mathbf{A_z'}$ is $\mathbf{P_c'}$ itself.
Figure \ref{fig:model_II} shows more simulation results, which include three layers of polygons, $\mathbf{P_c'}$, $\mathbf{P_v'}$ and $\mathbf{A_z'}$.

{\setlength{\tabcolsep}{0pt}
\begin{figure}[h]
\centering
\begin{tabular}{cc}
\includegraphics[height=.4\linewidth]{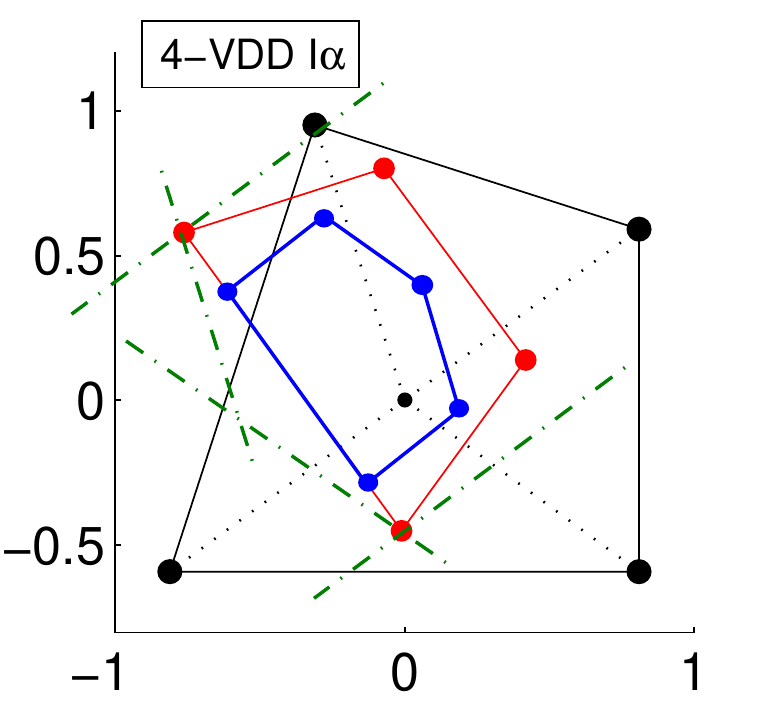}&
\includegraphics[height=.4\linewidth]{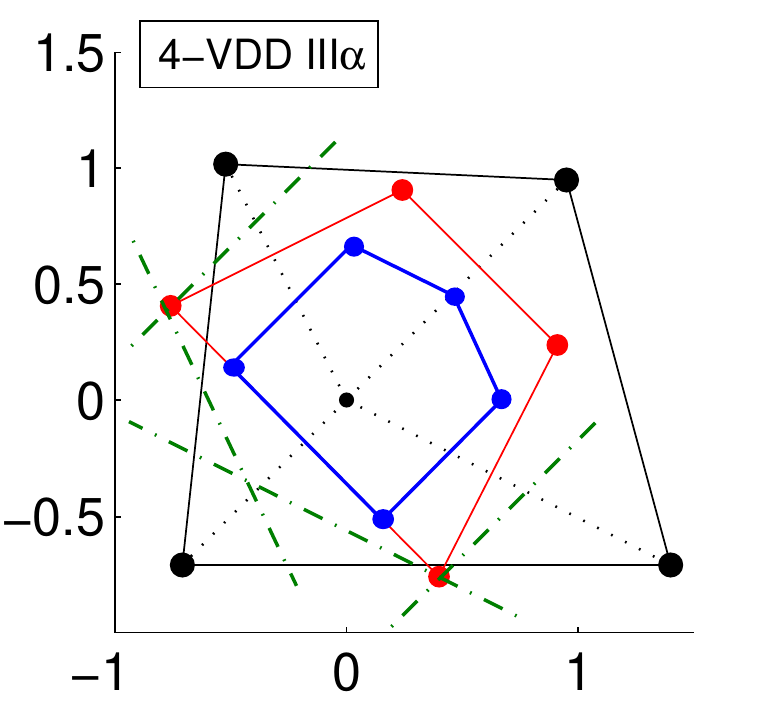} \end{tabular}
\caption{Sector shifting examples for $n=4$.
\label{fig:vdd_n4}}
\end{figure}
}

\subsection{Model III - Double Shifting}
Model III introduces the irregularity to both the exterior Delaunay polygon and interior Voronoi polygon. We first shift the exterior polygon to $\mathbf{P_c'}$ using Algorithm 1, then compute the VDD on $\mathbf{P_c'}$ to get the irregular Voronoi polygon $\mathbf{P_v'}$, and finally update $\mathbf{P_v'}$ by shifting using Algorithm 2. Figure \ref{fig:model_III} shows the simulation results, which include three layers of polygons, $\mathbf{P_c'}$, $\mathbf{P_v'}$ and $\mathbf{A_z'}$.

\subsection{Models $\alpha$, I$_\alpha$, II$_\alpha$, III$_\alpha$ - Sector Shifting}
Models I$_\alpha$, II$_\alpha$, III$_\alpha$ introduce another dimension of irregularity by shifting sector rays from the original equal divisions ($\alpha=\frac{2\pi}{n}$) while generating $n$ sectors. The surrounding angles are computed as follows:
\begin{equation}
\alpha_i \leftarrow (1\pm \epsilon)\alpha, \forall i=0, \ldots,n-1,
\end{equation}
where a random number $\epsilon=Rand[0,\kappa]$, and we select $\kappa\in \{0, 0.02, 0.04, 0.06, 0.08, 0.1\}$ in our experiments. The sector shifting (on sector rays) is not conflict with the previous interior/exterior shifting (on vertices/edges with fixed sector rays). Once we get the exterior polygon with random sector sifting, we then perform the algorithms of Models I-III to generate the result. Figures \ref{fig:model_I}, \ref{fig:model_II} and \ref{fig:model_III} show the simulation results, which include three layers of polygons, $\mathbf{P_c'}$, $\mathbf{P_v'}$ and $\mathbf{A_z'}$. For $n=4$, the interior polygon $\mathbf{P_c'}$ is not a rectangle because of the unequal sector angles and the perpendicular lines at vertices with acute corner angle are outside $\mathbf{P_c'}$, as shown in Fig. \ref{fig:vdd_n4}.

\subsection{Analysis}
The following guarantees that the obtained anonymity zone is qualified to protect user location from attacking.

\begin{lemma}
\label{lem:o_a_z}
The seed $O$ is within the anonymity zone.
\end{lemma}

\textsl{Proof:} Based on the definition of Voronoi-Delaunay duality,
the Voronoi edge $\overline{V_iV_{i+1}}$ is the perpendicular bisector of the sector ray $\overline{OC_{i+1}}$, and the Voronoi vertices $V_i,V_{i+1}$ are in the two sides of $\overline{OC_{i+1}}$, respectively. According to the construction of the anonymity zone, the seed $O$ is within the feasible strip $\mathbf{\Gamma}_i$, $O \in \Gamma_i$, 
and $\mathbf{A_z}=\cap \{\mathbf{\Gamma}_i\}$.
Thus $O \in \mathbf{A_z}$. Proof for the finally generated irregular $\mathbf{A_z'}$ is similar.

\begin{lemma}
The probability of any point in anonymity zone to be the seed $O$ is equally likely. 
\end{lemma}

\textsl{Proof:} According to Lemma \ref{lem:o_a_z}, given the anonymity zone $\mathbf{A_z}$, the user location $O \in \mathbf{A_z}$. Using the conclusions in attack analysis Section \ref{sub_sec:attack_analysis} for all the models, every point in the anonymity zone could be the user location.
In theory, if the sampling resolution of the anonymity zone goes to infinity, then the probability of revealing the exact location of $O$ goes to zero. In practice, if it is known that there are $M$ points in $\mathbf{A_z}$ for attempting, then the probability of identifying the user is $\frac{1}{M}$; if no such preconditions, the probability to reveal $O$ is infinitely  small (zero). With the same sampling resolution, the greater the area of anonymity zone $Area(\mathbf{A_z})$ (the greater $M$), the more difficult to reveal the seed.

\begin{lemma}
The anonymity zone area decreases with the increase of $n$, for the same radius of interest $r$.
\label{lem:Az_n}
\end{lemma}

\textsl{Proof:}
According to Lemma \ref{lem:o_a_z}, anonymity zone is the intersection region of all the lines drawn at Voronoi vertices and perpendicular to Voronoi edges. So, the number of the perpendicular lines are $2n$. With the increase of $n$, a convex polygon tends to be more circular, then the resulted (shifted) Voronoi edges become shorter, the perpendicular strips become narrower, and therefore, the area of the intersections of perpendicular strips (anonymity zone) decreases. This can also be observed in Fig. \ref{fig:comparison}.

\section{Performance Evaluation}
\label{sec:experiment}

 In this section, we evaluate the proposed $n$-VDD models in terms of the main principles: privacy level $\Gamma$, concealing cost $\Psi$, and communication cost $\Omega$. We demonstrate the efficiency and efficacy of the $n$-VDD models by experimental simulations and comparison.

\subsection{Experimental Setup}
 We first consider a network region, with a square area of $10^4 m \times 10^4 m$. The location of the user is randomly picked within this region. The user's region of interests (ROI) with a radius will always be within this network region. The different parameters used to evaluate the performance of the proposed methods include the number of vertices $n$, the range for random angle shifting $\kappa$, and the expected neighborhood radius around the user or of the region of interest $r$ (meter). We apply this parameter setting to test all the models. For each combination of these parameters, we perform 1000 iterations of the algorithm and then compute the average value of concealing cost and privacy level to generate the statistics.

\begin{figure}[t]
\centering
\footnotesize
\begin{tabular}{cc}
\includegraphics[height=0.42\linewidth]{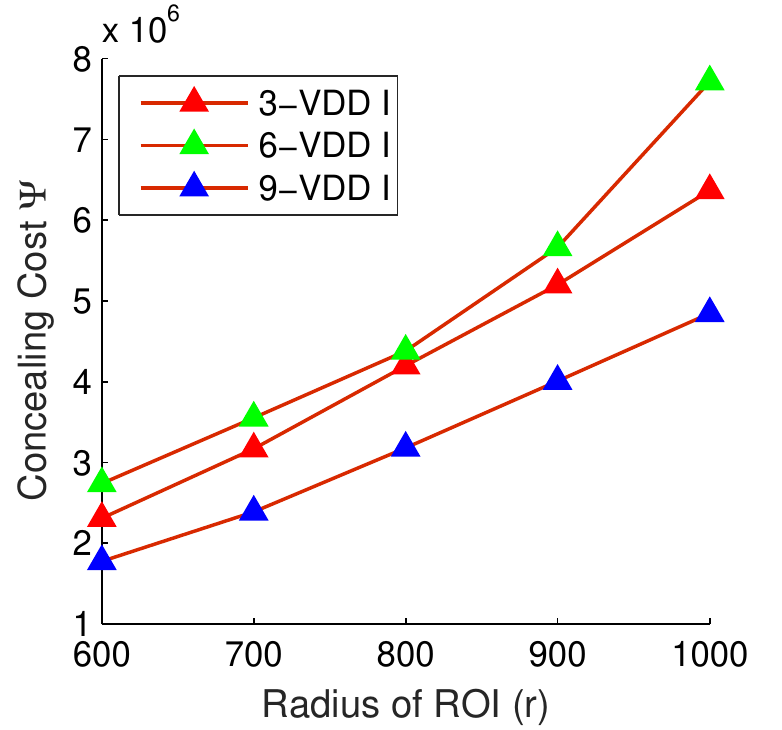}&
\includegraphics[height=0.42\linewidth]{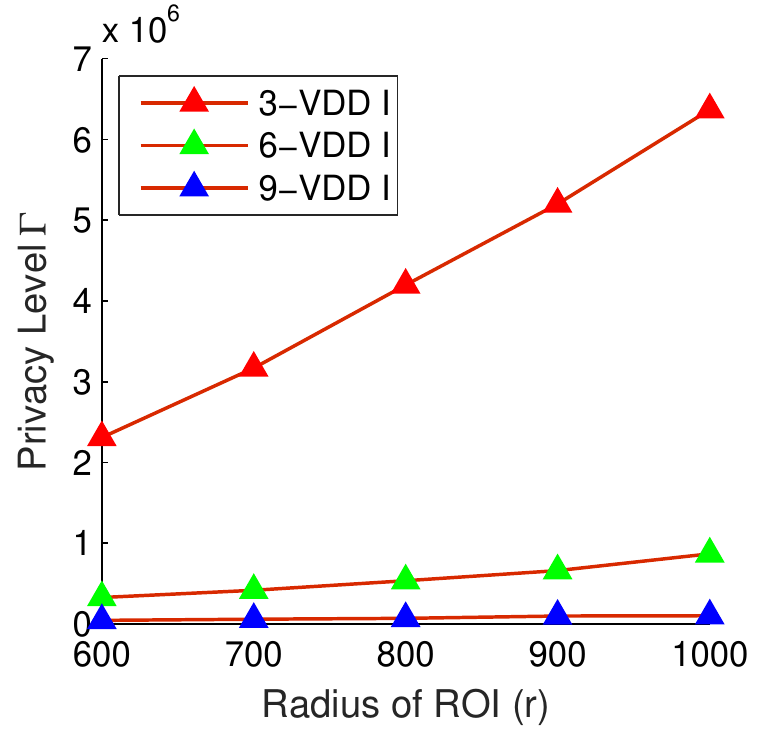}\\
\end{tabular}
\caption{Concealing cost $\Psi$ and privacy level $\Gamma$ of VDD models in terms of different $n$ and $r$. }\label{fig:vdd_concealing_privacy_r}
\end{figure}

\subsection{Concealing Cost, Privacy Level}
We define the concealing cost $\Psi$ as the measurement of the area of the concealing space, i.e., the scaled Voronoi polygon $\mathbf{P_v}^*$, for all the models.
We then define the privacy level $\Gamma$ as the measurement of the area of the anonymity zone, i.e., the scaled most interior polygon $\mathbf{A_z}^*$.

\begin{figure}[t]
\centering
\footnotesize
\begin{tabular}{cc}
\includegraphics[height=.42\linewidth]{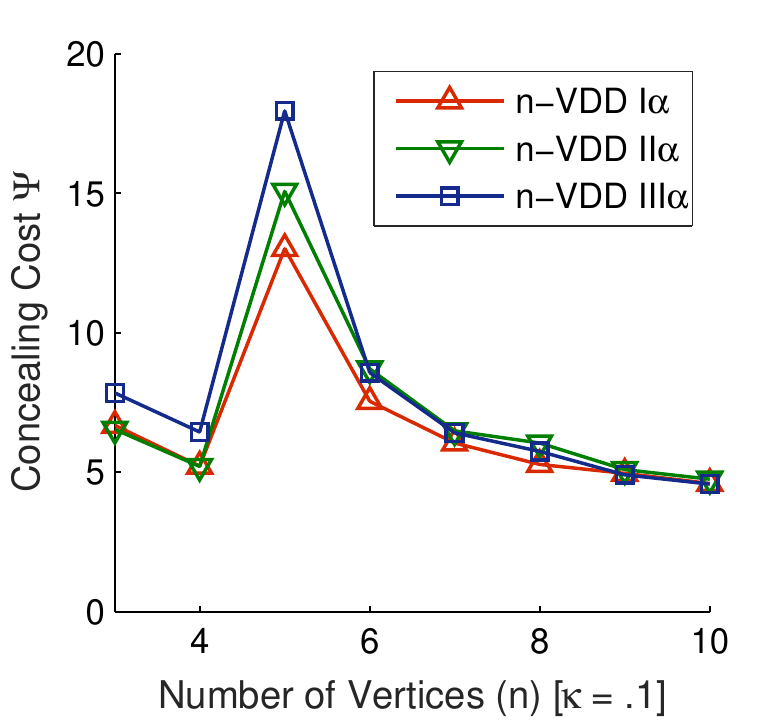}&
\includegraphics[height=.42\linewidth]{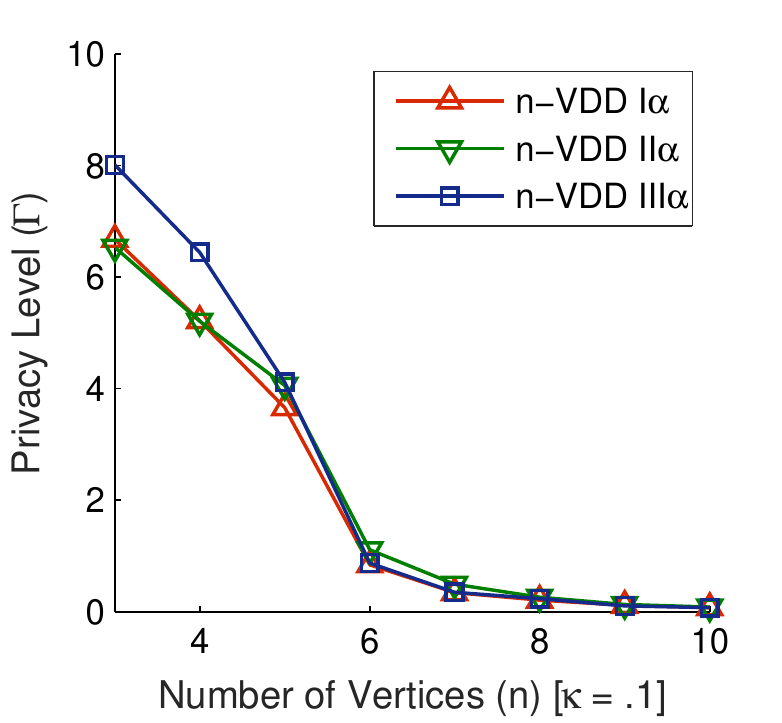}\\
\end{tabular}
\caption{Concealing cost $\Psi$ and privacy level $\Gamma$ of VDD models in terms of different $n$ ($r=1.0,\kappa=0.1$).\label{fig:n}}
\end{figure}

\begin{figure}[t]
\centering
\footnotesize
\begin{tabular}{cc}
\includegraphics[height=.42\linewidth]{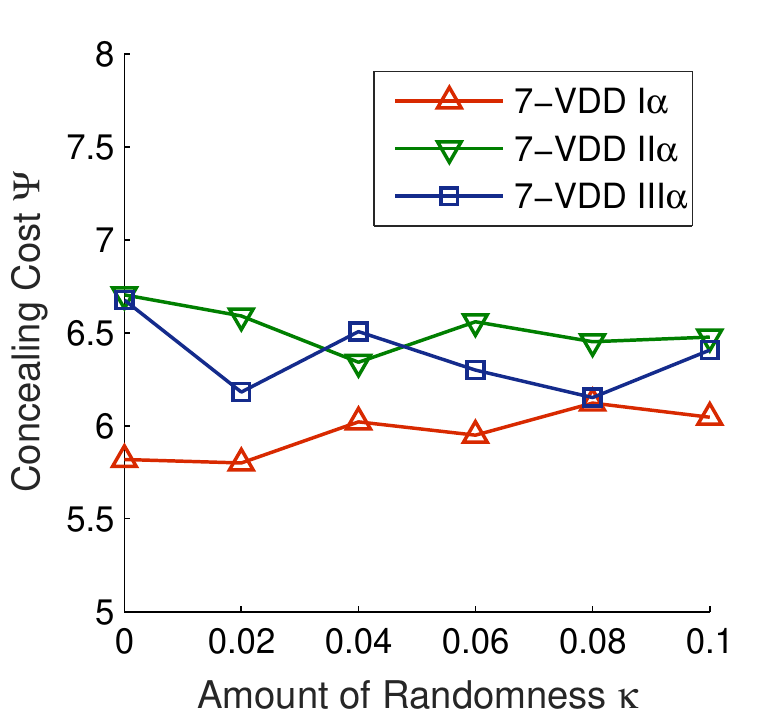}&
\includegraphics[height=.42\linewidth]{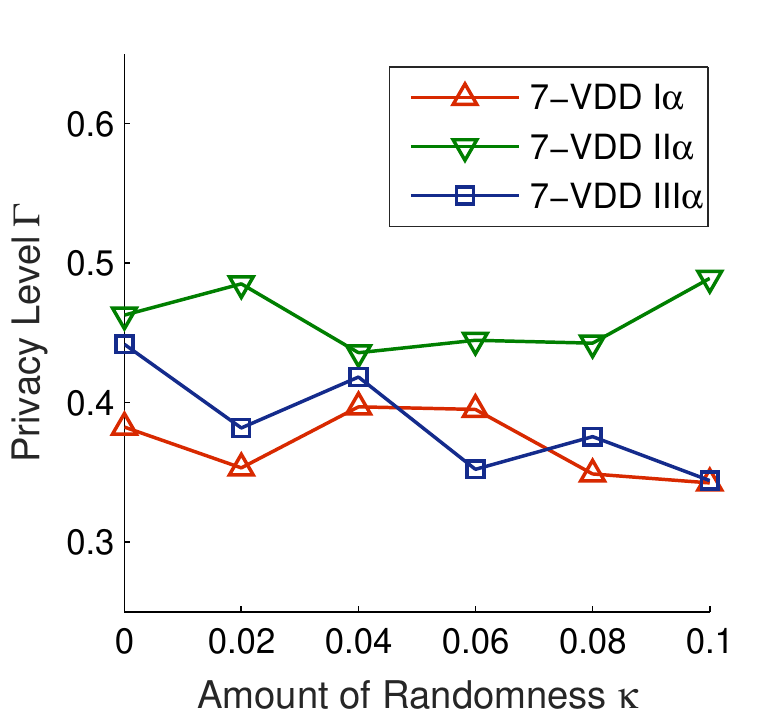}\\
\end{tabular}
\caption{Concealing cost $\Psi$ and privacy level $\Gamma$ of VDD models in terms of different angle randomness range $\kappa$ ($r=1.0, n=7$).\label{fig:k}}
\end{figure}

\subsubsection{Impact of Radius of ROI}
In theory, with the increase of the radius $r$ of ROI, the concealing cost $\Psi$ and the privacy level $\Gamma$ will linearly increase. This can be easily explained in our method as follows: in our settings, we set the initial radius as $r_0=1.0$. Then after generating the shifted Voronoi polygon $\mathbf{P_v}'$, we adapt it to get the concealing space by a linear scaling with scalar $\lambda$ (linear to the customized $r$) to cover the ROI. Therefore, we have the linear theoretic claim.
As shown in Fig. \ref{fig:vdd_concealing_privacy_r}, an almost similar upward linear trend of $\Psi$ and $\Gamma$ appears for different $n$ over $r$ (Model I). Other $n$-VDD models have similar results.

\subsubsection{Impact of Number of Vertices $n$}
We analyze the concealing cost $\Psi$ and the privacy level $\Gamma$ with different $n$ for all the models. As shown in Fig. \ref{fig:n}, $\Psi$ decreases when $n$ changes from 3 to 5, but increases when $n$ increases above 5, while $\Gamma$ decreases with the increase of $n$ from 3 (Model I). The plots for other models are similar in our simulations. This is consistent to Lemma \ref{lem:Az_n}.

\subsubsection{Impact of Angle Randomness Range $\kappa$}
We analyze the concealing cost $\Psi$ and the privacy level $\Gamma$ with different range of angle randomness $\kappa$ for all the models. As shown in Fig. \ref{fig:k}, the values are almost linear for different $\kappa$.

\subsection{Communication Cost}
We compute the communication cost $\Omega$ as the sum of upstream and downstream cost in the network traffic.

\subsubsection{Upstream Traffic} For the $n$-sided concealing polygon $\mathbf{P_v}^*$, the upstream traffic is computed as $48 + 8 \times n$, where a packet header costs 40 bytes, 8 bytes are added for the user ID $u_{id}$,
and then 8 bytes are used to represent the $x,y$ coordinates of each ${V_i}^*$.

\subsubsection{Downstream Traffic}
For the $N$ number of preferences of interest (POIs) returned, the downstream traffic is computed as
$40 + 8 \times N$, where a packet header costs 40 bytes,  and each POI $x,y$ coordinates costs 8 bytes.

\subsection{Comparison}
In this work, we compare our method with the closely related method, the $n$-CD model \cite{6567113}, which uses $n$ concealing disks to cover the user location and the region of interest, reporting the rotated disk centers and radii of the LBS server. The intersection region of all these concealing disks defines the anonymity zone, as shown in Fig. \ref{fig:nCD}. In terms of computation, the proposed $n$-VDD models are easy to implement and practical; the computations are mainly based on line-line intersections, and therefore the anonymizing algorithm is linear. In contrast, the $n$-CD model requires circle-circle intersections which is computationally more expensive. Thus, our method is more efficient than the $n$-CD one. The following details the comparison of the anonymizing performance in terms of concealing cost, privacy level, and communication cost (see the plots in Fig.  \ref{fig:comparison} for different $n$ with $r=10^3, \kappa=0.02$).

{\setlength{\tabcolsep}{0pt}
\begin{figure*}[t]
\centering
\footnotesize
\begin{tabular}{ccc}
\includegraphics[width=.32\linewidth]{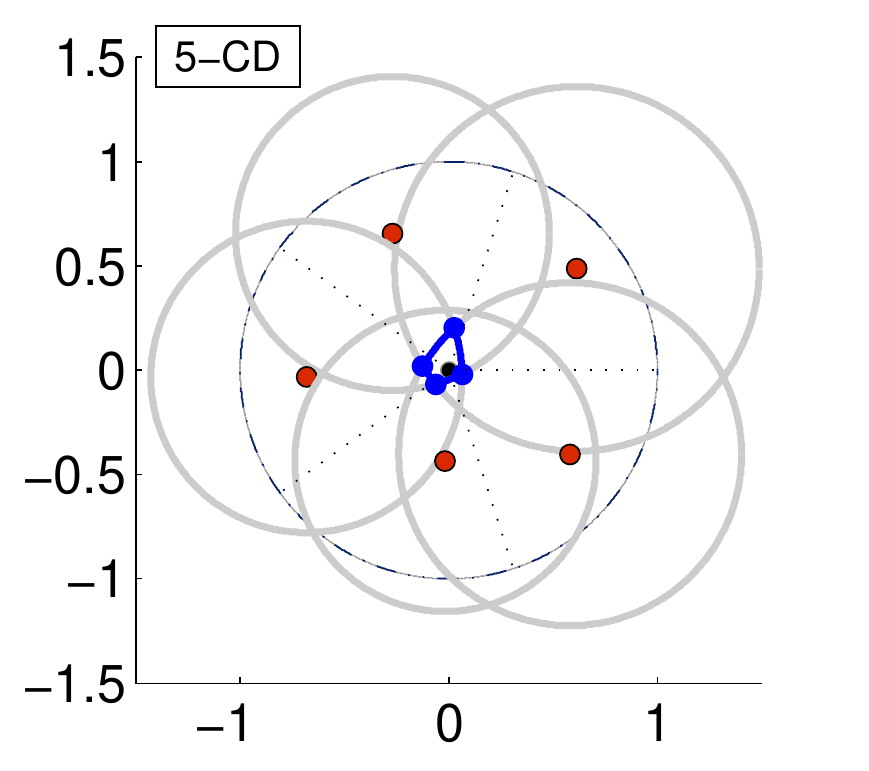} &
\includegraphics[width=.32\linewidth]{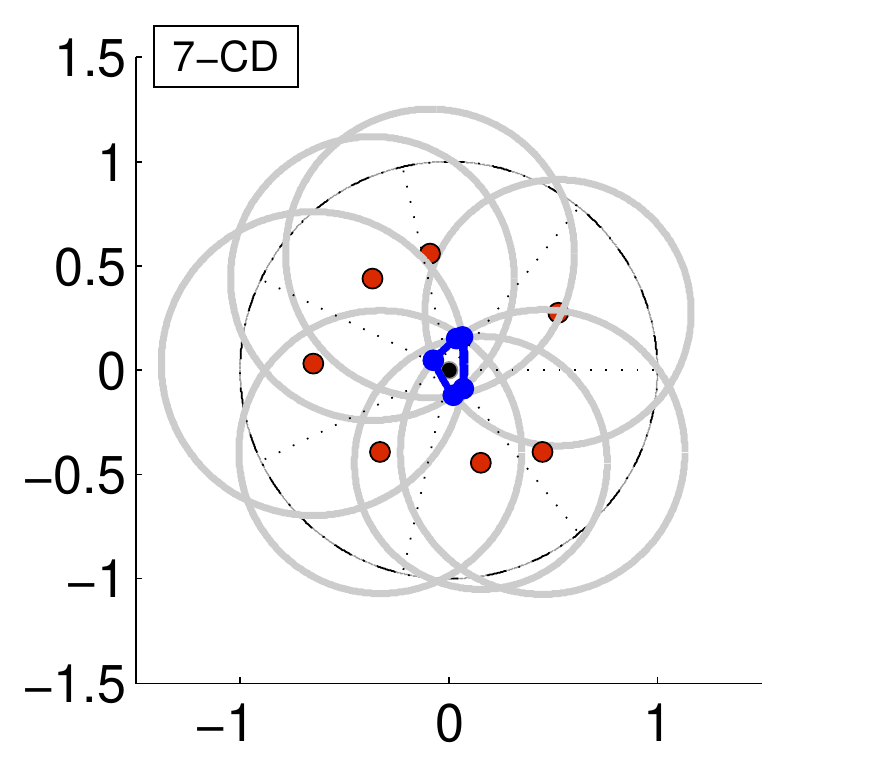} &
\includegraphics[width=.32\linewidth]{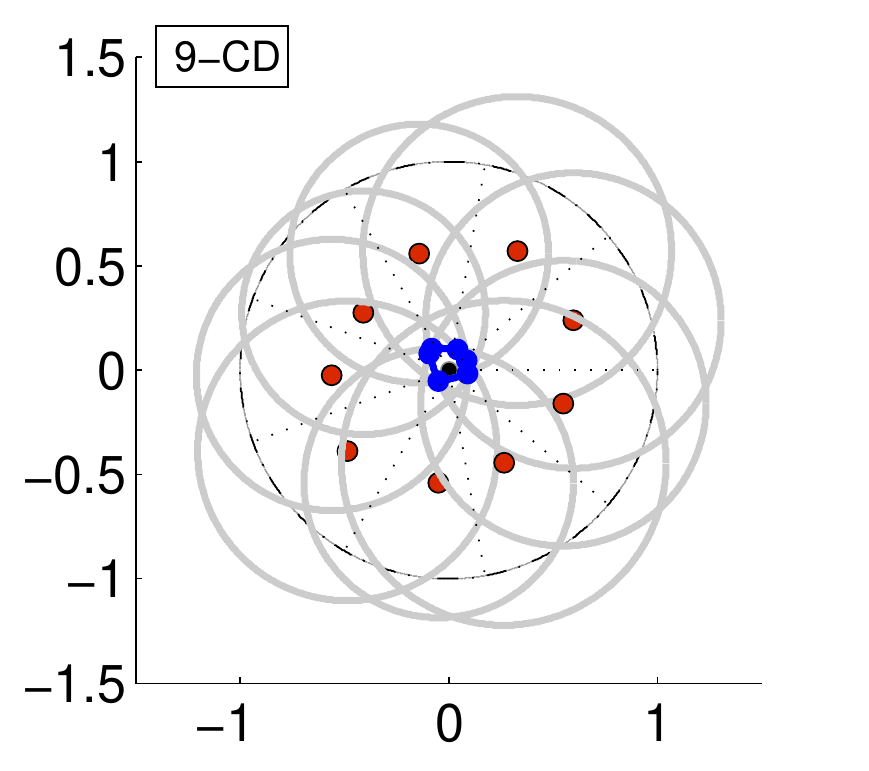} \\
\end{tabular}
\caption{Simulation examples of the $n$-CD model \cite{6567113} for comparison ($r=1.0$). \label{fig:nCD}}
\end{figure*}
}
{\setlength{\tabcolsep}{0pt}
\begin{figure*}[t]
\centering
\footnotesize
\begin{tabular}{ccc}
\includegraphics[height=.26\linewidth]{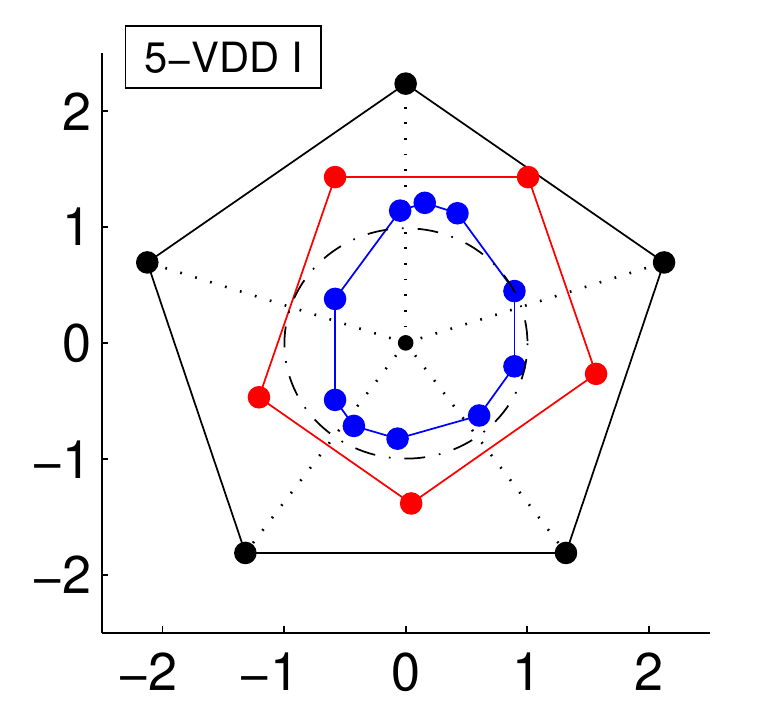} &
\includegraphics[height=.26\linewidth]{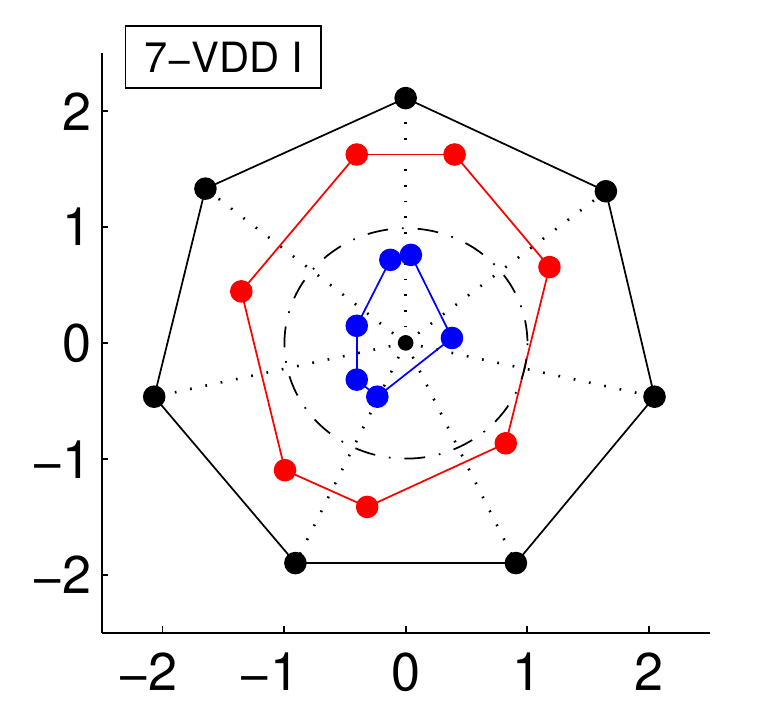} &
\includegraphics[height=.26\linewidth]{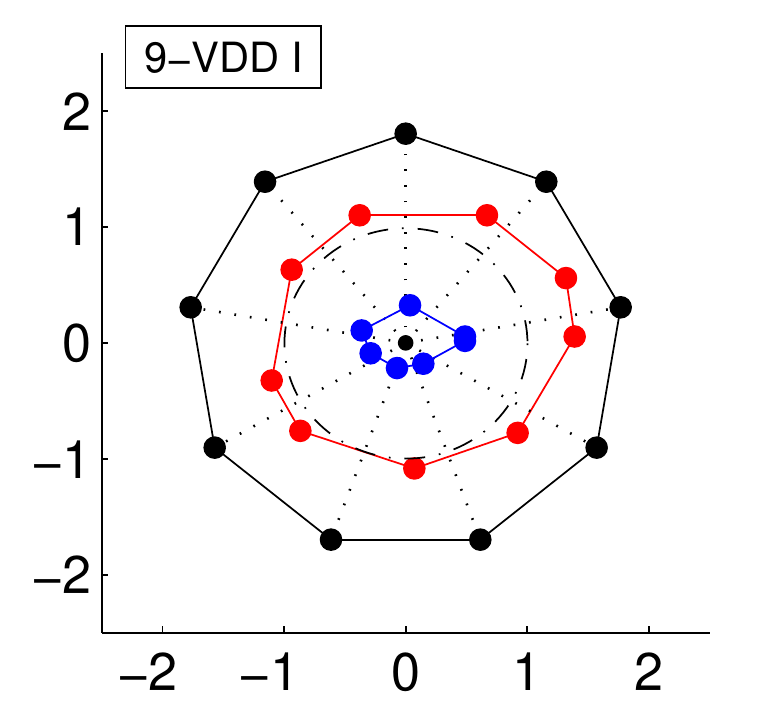} \\
\end{tabular}
\caption{Simulation examples of the $n$-VDD model I after scaling ($r=1.0$). \label{fig:nCD}}
\end{figure*}
}

{\setlength{\tabcolsep}{0pt}
\begin{figure*}[t]
\centering
\small
\begin{tabular}{ccc}
\includegraphics[height=.3\linewidth]{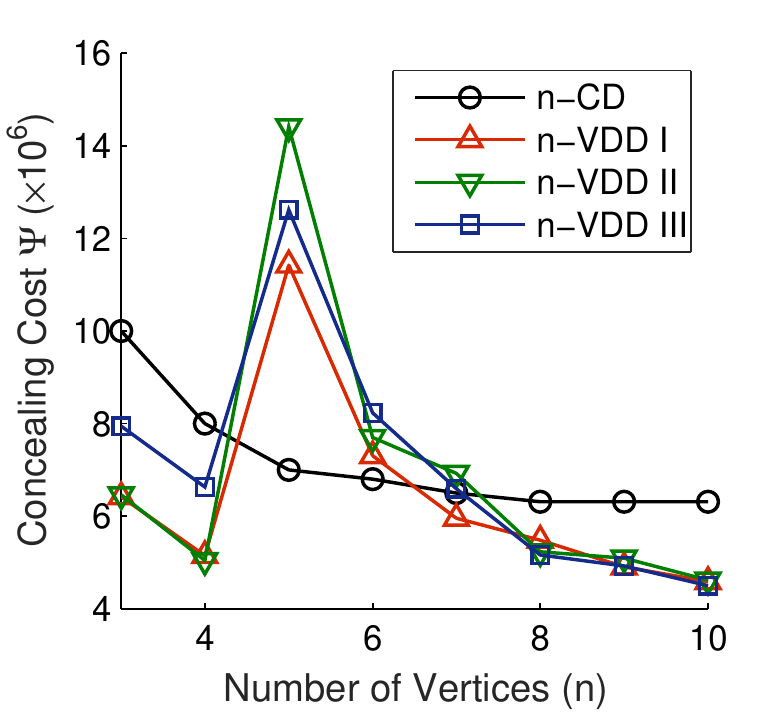}&
\includegraphics[height=.3\linewidth]{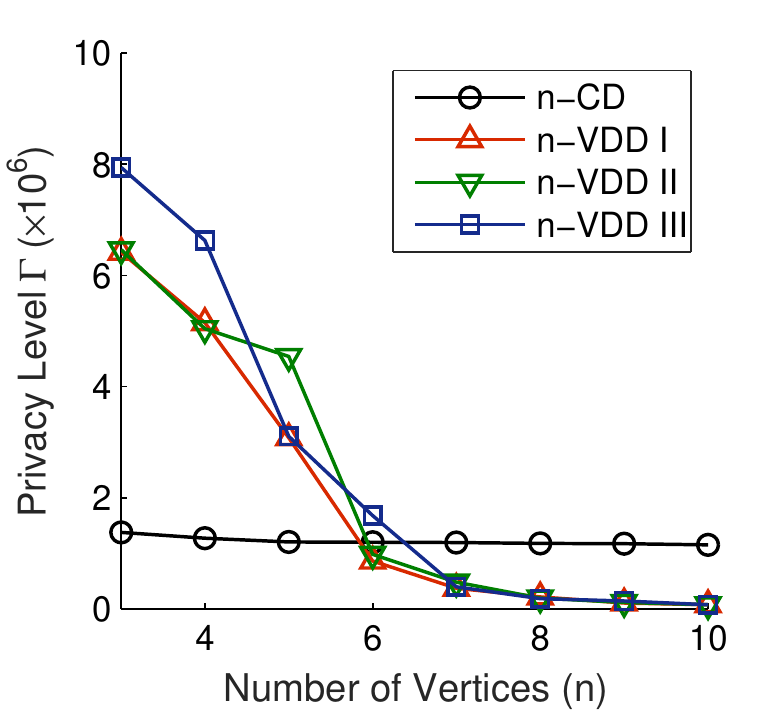}&
\includegraphics[height=.3\linewidth]{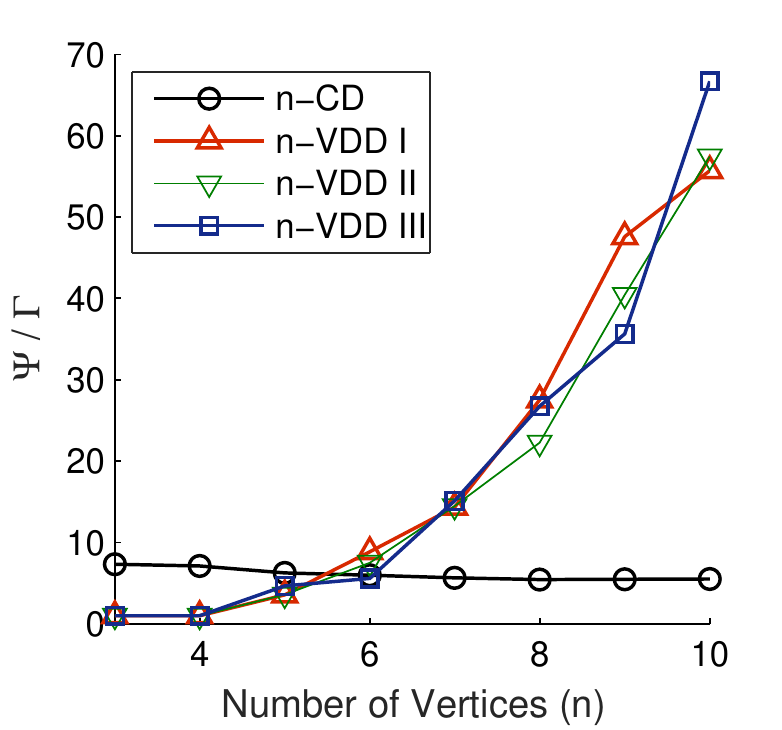}\\
(a) Concealing cost $\Psi$ & (b) Privacy level $\Gamma$ & (c) Ratio $\Psi/\Gamma$\\
\end{tabular}
\caption{Comparison between the $n$-VDD models and $n$-CD model in terms of $n$ ($r=10^3$).}\label{fig:comparison}
\end{figure*}
}

\subsubsection{Concealing Cost, Privacy Level}
Figure \ref{fig:comparison}(a-b) shows the plots for the $n$-VDD models I-III, in which, for $n=3,4$, the concealing space and the anonymity zone overlap.
We observe that in the $n$-VDD models, the concealing cost has a peak at $n=5$ and decreases from $n=5$, and the privacy level decreases as the value of $n$ increases; in the $n$-CD model, they are almost linear (almost identical). While the number of vertices $n$ is small ($n=3,4$) and big ($n\geq 8$), the concealing cost of the $n$-VDD models are much lower than $n$-CD when the radius of ROI $r$ ($10^3$) is fixed.
We can simply get that one can achieve higher privacy level with lower concealing cost using the models of $n$-VDD than $n$-CD with smaller value of $n$.
Minimizing the trade-off between $\Psi$ and $\Gamma$ is the primary goal of an anonymizing protocol. We analyze and compare this trade-off as the ratio of the two terms $\Psi$, $\Gamma$ for the $n$-CD and the $n$-VDD. Figure  \ref{fig:comparison}(c) shows that for smaller value of $n$ ($\leq 5$), this ratio in the models of $n$-VDD is much smaller than $n$-CD, i.e., we can have higher privacy level compared to concealing cost for smaller value of $n\leq 5$ than $n$-CD. At $n = 3,4$, the ratio in $n$-VDD is 1.0 because the concealing space and the anonymity zone overlap, while $n$-CD is much higher. At $n=6$, the two models have close values. Therefore, we can select $n\leq6$ in practice. This trade-off shows the strength of our proposed methods in case of smaller values of $n$. In the other hand, there is flexibility to adapt the values (performance) by changing $n$ in our models, while the values for the $n$-CD model are almost stable with different $n$.

\subsubsection{Communication Cost}

In the $n$-CD model, the upstream traffic cost is calculated as $48 + 12 \times n$, where $n$ denotes the number of concealing disks, $48$ is the sum of 40 bytes for the packet header and 8 bytes for the user id, $12$ is the sum of 8 bytes for disk center $x, y$ coordinates and 4 bytes for disk radius; the downstream traffic cost is calculated as $40 + 8 \times N$, where $N$ denotes the number of the returned POIs and each has $8$ bytes for $x,y$ coordinates.
In terms of upstream cost, we observe that the $n$-VDD models are better than $n$-CD. The downstream cost largely depends on the area of the concealing space: the larger the concealing cost, the more the downstream cost. Moreover, if the concealing cost is high, the quality of services will be low. So we can define both downstream cost and quality of services as functions of concealing cost. From Fig.  \ref{fig:comparison}(c), we see that the concealing cost of the $n$-VDD models are lower than that of $n$-CD, implying the maximum bound of both the communication cost and quality of services error are lower.

\section{Conclusion}
\label{sec:conclusion}

We present a novel location privacy framework, the so-called $n$-VDD, based on the Voronoi-Delaunay duality (VDD). This work is based on the insight that only an irregular Voronoi cell around the user location can not induce the user location (or the seed) but can give the anonymity zone, which is the intersection of all the parallel strips perpendicular to and bounded by Voronoi edges. We introduce the irregularity to the Voronoi cell using three terms for random shifting and their combinations: (1) interior shifting - starting from a regular polygon and shifting the generated Voronoi cell to be irregular with randomness, (2) exterior shifting - starting from an irregular polygon generated with randomness, and (3) sector shifting - starting from a regular polygon and shifting the sector rays around the seed.
All the computations are efficient based on basic planar geometry and linear algebra. Experiments and comparisons have demonstrated the efficiency and efficacy for protecting the user location.
In future work, we will explore the location privacy applications in 3D environment by generalizing planar $n$-VDD to volumetric $n$-VDD.


%


\section*{Acknowledgment}
This work was supported by NSF CCF-1544267, NSF CNS-1263124/15601334 and NSF CNS-1407067.


\ifCLASSOPTIONcaptionsoff
  \newpage
\fi



%
\bibliographystyle{plain}
\end{document}